\documentclass{emulateapj}

\shortauthors{Matthews}
\shorttitle{Radio Stars Workshop Summary}

\begin{document}
\newcommand{\ang}{\rm \AA}
\newcommand{\msun}{M$_\odot$}
\newcommand{\lsun}{L$_\odot$}
\newcommand{\days}{$d$}
\newcommand{\degree}{$^\circ$}
\newcommand{\ud}{{\rm d}}
\newcommand{\as}[2]{$#1''\,\hspace{-1.7mm}.\hspace{.0mm}#2$}
\newcommand{\am}[2]{$#1'\,\hspace{-1.7mm}.\hspace{.0mm}#2$}
\newcommand{\ad}[2]{$#1^{\circ}\,\hspace{-1.7mm}.\hspace{.0mm}#2$}
\newcommand{\lsim}{~\rlap{$<$}{\lower 1.0ex\hbox{$\sim$}}}
\newcommand{\gsim}{~\rlap{$>$}{\lower 1.0ex\hbox{$\sim$}}}
\newcommand{\HA}{H$\alpha$}
\newcommand{\HII}{\mbox{H\,{\sc ii}}}
\newcommand{\kms}{\mbox{km s$^{-1}$}}
\newcommand{\HI}{\mbox{H\,{\sc i}}}
\newcommand{\KI}{\mbox{K\,{\sc i}}}
\newcommand{\nan}{Nan\c{c}ay}
\newcommand{\jks}{Jy~km~s$^{-1}$}
\slugcomment{Accepted to PASP}

\title{Radio Stars and Their Lives in the Galaxy}

\author{Lynn D. Matthews\altaffilmark{1}}
\altaffiltext{1}{MIT Haystack Observatory, Off Route 40, Westford, MA
  01886 USA; lmatthew@haystack.mit.edu}

\begin{abstract}

This paper summarizes the
three-day international workshop {\it Radio Stars and Their Lives
in the Galaxy}, held  at the Massachusetts Institute of Technology
Haystack Observatory on 2012 October 3-5. The
workshop was organized to provide a forum for the presentation and discussion of
advances in stellar and solar astrophysics recently (or soon
to be) enabled by the latest generation of state-of-the-art observational
facilities operating from meter to submillimeter wavelengths. The
meeting brought together
both observers and theorists to discuss  
how radio wavelength observations are
providing new and unique insights into the workings of stars and their
role in the Galactic ecosystem. Topics covered 
included radio emission from hot and cool stars (from the pre- to
post-main-sequence), the Sun as a radio star, circumstellar chemistry,
planetary nebulae, white
dwarf binaries and novae, supernova progenitors, and radio stars as
probes of the Galaxy. 

\end{abstract}


\section{Background and Motivation for the Workshop}
Radio emission has now been detected from stars across the entire
Hertzsprung-Russell (H-R) diagram, spanning virtually every stage of
stellar evolution. 
The recent commissioning of a host of new
and upgraded observational facilities operating at wavelengths
spanning from the meter to the submillimeter has been leading to a
rapid pace of discoveries in nearly every branch of stellar
astrophysics.
Motivated by these advances, a 3-day workshop entitled ``Radio Stars
and Their Lives in the Galaxy'' was convened at the Massachusetts
Institute of Technology (MIT) Haystack Observatory in Westford,
Massachusetts on 2012 October 3-5. The workshop drew a total of 50 participants from four
continents and 13 countries and brought together both
theorists and observers to participate in a program of oral
presentations, posters, and discussions. The scientific program was
devised to
bring together stellar astrophysicists from a variety of sub-disciplines to stimulate
cross-fertilization of ideas on unsolved problems common to the study of many
classes of stellar objects
(e.g., mass loss, the driving of bipolar outflows, and the generation of magnetic fields).

The first international workshop on Radio Stars took place in
Ottawa, Canada in June 1979, at which time the discipline of stellar
radio astronomy was approximately a decade old (Feldman \& Kwok
1979). 
The Very Large Array (VLA) had not yet been commissioned, hence available data were
largely limited to single-dish flux density measurements at
centimetric wavelengths without the advantages of imaging. 
Thirty-three years later, the Haystack
workshop made abundantly clear that stellar radio
astronomy remains a vibrant field that is witnessing a rapid pace of advances
relevant to many of the most
important topics in stellar astrophysics. We witnessed how new technologies
are bringing dramatic advances in areas 
such as resolved imaging of radio photospheres, molecular line imaging, 
dynamic imaging
spectroscopy, high-precision astrometry, and the routine study of
sources at a level of a few tens of $\mu$Jy. We
also heard how developments in theory, modeling, and laboratory
astrophysics are improving our ability to interpret results. This paper attempts to
summarize a sampling of the diverse array
topics presented and discussed at the Haystack workshop. 

\section{Scientific Sessions}
The science program of the Haystack Radio Stars workshop comprised 
33 oral presentations (13
invited reviews and 20 contributed talks) and 7
posters. The complete program and presentation abstracts
are available at http://www.haystack.mit.edu/workshop/Radio-Stars/.

The workshop was organized into seven scientific sessions: (I) An
Overview of Stellar Radio Astronomy; (II) Radio Emission from Pre-Main
Sequence Stars; (III) Radio Emission on the Main Sequence; (IV) 
Post-Main Sequence Radio Stars;
(V) Stellar Remnants; (VI) Radio Stars as Players in our Galaxy;
(VII) Summary and Perspectives. This summary paper loosely
follows this organization, but also attempts to synthesize 
recurrent themes and topics that spanned multiple sessions. 
In addition to summarizing new results presented by participants, I
strive to draw attention to topics where there was a consensus
that our
understanding is still limited, where controversies persist,  
and/or where future investigations are likely to
prove fruitful. 
I close with a summary of
concerns and future challenges raised during the final discussion session.

\section{An Overview of Stellar Radio Astronomy}
M. G\"udel (University of Vienna) opened the workshop with a comprehensive overview of
the versatility of the radio spectrum for understanding physical
processes in stars across the entire H-R diagram. He
reviewed radio
emission mechanisms important for stellar sources and emphasized that
magnetic fields have now been shown to be important in nearly every type of stellar radio
source. This point was underscored in many subsequent talks. 
G\"udel cited several ``keywords'' expected to define the future of stellar radio astronomy,
including {\em $\mu$Jy sensitivity}, {\em time resolution,
bandwidth, spectral resolution}, and {\em frequency
coverage}. S. Kwok (The University of Hong Kong) reminded us of
another during his closing talk: {\em dynamic range}. 
Thanks to the truly impressive suite of new and upgraded
radio facilities that has come on-line during the past few years,
the workshop provided no shortage of illustrations of how recent
gains in each of these key areas are leading to
breakthroughs in our understanding of the workings of stars.

\section{Radio Emission from Pre-Main Sequence Stars}
\subsection{The Radio-X-ray Connection in Young Stellar Objects\protect\label{ysos1}}
It has been known for some time that young stellar objects (YSOs) are producers of both
X-ray and centimetric radio emission. However, 
as summarized by S. Wolk (Harvard-Smithsonian Center for
Astrophysics), to date
X-ray and radio observations of
young clusters have revealed a paltry number of simultaneous
detections, with only a few percent of YSOs found to show both types of
emission (Forbrich et al. 2011; Forbrich \& Wolk 2013). This comes as a surprise, since many
of the YSOs are predicted to be bright enough for detection in the
radio based on the G\"udel-Benz relation\footnote{The G\"udel-Benz
  relation  (G\"udel \& Benz 1993; Benz \& G\"udel 1994)
is a tight empirical correlation between the quiescent X-ray
luminosity and 
radio luminosity for active stars that spans more than 10 orders of
magnitude. This relationship suggests an intimate connection
between the non-thermal electrons responsible for the radio emission
and the thermal plasma giving rise to the X-rays.},
yet fall short by two orders
of magnitude or more. At the same time, several of
the YSOs that are detected in the radio lie  {\em above} the G\"udel-Benz
relation. Further, the radio detections are not correlated with any
identifiable X-ray properties of the sources. Together these results suggest that 
in YSOs the X-ray and radio emission are
decoupled. 

As described by Wolk, the Jansky Very Large Array (JVLA) is soon expected to
dramatically extend progress in YSO studies, providing a jump by a
factor of $\sim$20 in sensitivity over past VLA studies. This will greatly improve
radio detection statistics and allow polarization measurements for many
more sources. In addition, the JVLA will provide the spectral coverage critical
for obtaining radio spectral indices and thus distinguishing between
thermal and non-thermal emission mechanisms. Wolk described an ongoing campaign to
simultaneously observe regions of Orion with {\it Chandra} and the JVLA, with
first results expected in late 2012. 

\subsection{Radio Emission from the Jets of Young Stellar Objects}
In addition to their
disks, winds, and coronae, the jets from YSOs are also a source of
centimetric radio emission. Dust-penetrating radio observations of
these jets provide 
the best available tool for addressing the still poorly understood problem of
their launch and collimation.
R. Ainsworth (Dublin Institute for Advanced Studies) presented
a 16~GHz continuum study of a sample of 16 
Class~0 and Class~I YSOs 
using the Arcminute Microkelvin Imager. Spectral indices derived
from a combination of the new measurements and other radio/sub-mm 
measurements from the literature point to thermal
bremsstrahlung as the dominant jet emission mechanism.  This
work also established that dust emissivity is important even at
centimeter wavelengths and that uncertainties in the dust opacity index comprise
the dominant uncertainty in calculations of the luminosity of
YSOs from radio data (Ainsworth et al. 2012).

\section{Radio Emission from Cool Main Sequence Stars}
\subsection{Emission from Ultracool Dwarfs\protect\label{ultracool}}
The interiors of stars
later than $\sim$M3 are thought to be fully convective, predicting
a breakdown in the $\alpha\Omega$ dynamo mechanism
for generating stellar magnetic fields. Furthermore,
at the ultracool end of the H-R diagram ($\gsim$M7), X-ray and H$\alpha$
emission are seen to dramatically decline relative to the stellar
bolometric luminosity, indicating sharp drops in chromospheric and
coronal heating. Consequently,
perhaps one of the greatest surprises in stellar radio astronomy in
the past decade has been the
discovery of radio emission from ultracool (late M, L, and T) dwarfs well in excess of 
what is predicted from the G\"udel-Benz relation, 
together with indications of very strong magnetic fields in some
cases.

\begin{figure*}
\centering
\scalebox{0.8}{\rotatebox{0}{\includegraphics{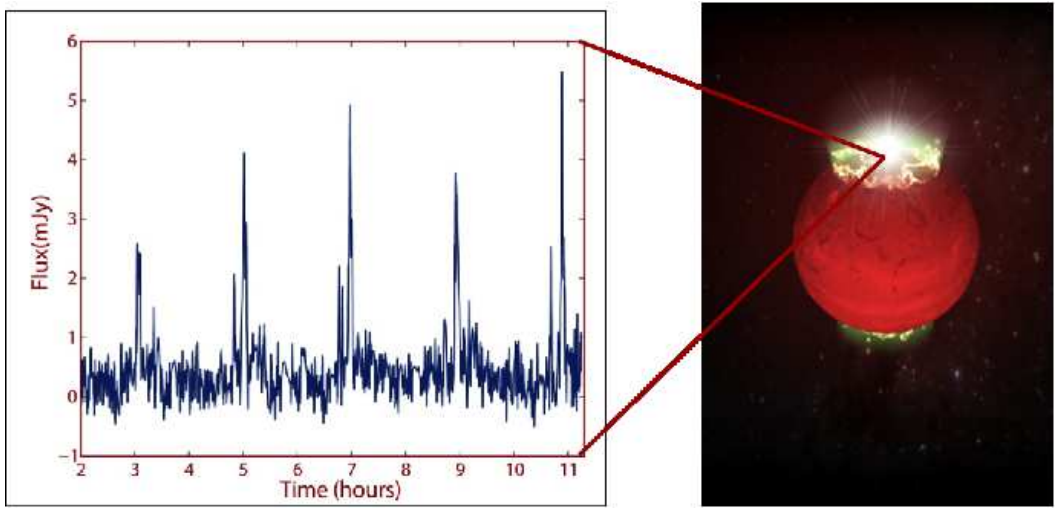}}}
\caption{
{\it Left:} An 8.44~GHz VLA light curve of the total
  intensity (Stokes~I) emission from the M9 dwarf
  TVLM~513-46546 (Hallinan et al. 2007). Periodic bursts of
coherent emission are evident. {\it Right:} Artist's conception of the 
radio-emitting region from an
  ultracool dwarf (courtesy of G. Hallinan). }
\label{fig:ultracool}
\end{figure*}

As described in the review by G. Hallinan (California Institute of Technology),
the emission from ultracool dwarfs  has been 
found to comprise both broadband, quiescent emission and 
periodic pulse-like flares. These flares
are nearly 100\% circularly polarized  and are 
rotationally modulated, analogous to beamed radiation
from pulsars. This points to
an electron cyclotron maser as the most
likely emission mechanism---the same mechanism responsible for the radio
emission from the planets in our solar system (e.g., Hallinan et
al. 2007, 2008; Figure~\ref{fig:ultracool}).
Importantly, this means that
the observed frequency cutoff of the emission can be directly
translated into a measure of the magnetic field strength, with values
in the kilogauss range reported for the  late M dwarfs and brown dwarfs studied to
date. 

Rotation increases rapidly from the late M through the T dwarfs, and
the correlation between radio
luminosity and rotation rate was noted by
several speakers. M.
G\"udel drew attention to a recent survey of $\sim$100 late-M and L dwarfs by  
McLean et al. (2012) that found that for spectral types $\gsim$M7, the
ratio of radio to bolometric luminosity increases with rotational velocity
without saturating. This result underscores that rotation controls
the magnetic activity and the 
production of the high energy electrons responsible for the radio
emission in these cool stars. However, it is not yet clear
whether rotation is powering the activity or the electron
acceleration. G\"udel suggested that a possible explanation for why
saturation does not occur at radio wavelengths  is that there are changes in surface
effects and or
filling factors in the latest types stars. For example, magnetic field
configurations that trap electrons and keep them levitated off the surface of
the star would force them to lose their energy via synchrotron
emission rather than colliding with other particles and
thermalizing. At the same time, low densities would explain why there is not much
X-ray emission.  On a related note, S. Cranmer (Harvard-Smithsonian Center for Astrophysics)
suggested that the combination of X-ray
and radio observations might help to establish whether there is
truly a basal ``floor'' to the age-rotation-activity relationship. 

Hallinan and G\"udel both highlighted the significance of
the recent detection of 4.75~GHz radio emission from a T6.5 dwarf by
Route \& Wolszczan (2012) using Arecibo. At $T_{\rm eff}\approx900$~K,
this star is
$\sim$1000~K cooler than any other dwarf detected in the radio to date. Its derived
field strength of $\sim$1.7~kG poses severe challenges for existing
dynamo theories for magnetic field generation. 
Radio measurements offer the only means of quantifying magnetic field
strength in this regime between stars and planets, since other traditional
techniques (e.g., exploiting the Zeeman effect) break down as molecular lines
become too rotationally broadened to yield useful measurements.

Despite the weakness of the optical and H$\alpha$ emission from
ultracool dwarfs,
Hallinan noted that in all cases where pulsed emission has been
detected, periodic variability in the broadband optical light and H$\alpha$
emission is also seen. He proposed that the periodic emission across
the entire spectrum from optical to radio is auroral in nature. As a test of this idea,
Hallinan cited results from a new program aimed at obtaining simultaneous
JVLA dynamic spectra and Keck optical spectrophotometry.
The JVLA dynamic spectra of three sources observed to date
permit derivation of the magnetic field strength as a function of
height, effectively mapping the auroral oval of the
sources. Correlated, periodic variability is also present across the
optical spectrum, arising from photospheric and chromospheric hot spots. 
The nature of the activity revealed in these objects
can be understood as being produced from large-scale, quasi-stable currents
in the magnetic fields rather than as a result of traditional coronal
activity. Comparison of these data with new models (e.g., Kuznetsov et
al. 2012) is expected to permit discrimination between
satellite-induced emissions versus rotationally modulated
emissions.

In spite of rapid growth in the study of ultracool dwarfs at radio
wavelengths, G\"udel and Hallinan both cautioned that only a minority of
cool dwarfs have so far been detected in the radio, and larger surveys
of cooler stars are needed for improving 
statistics and for gaining insight into the question of why only a
fraction of brown dwarfs are active. 
Another open question is how the electrons are accelerated in
ultracool dwarfs. Hallinan predicted that  lessons from planetary
magnetospheric physics may hold answers.
Finally, the nature of the {\em quiescent} emission from ultracool dwarfs (which
shows some modest variability) is still
poorly understood. Gyrosynchrotron emission is one candidate. Its
relation to the pulsed emission is also unknown.

\subsection{Radio Detection of Exoplanets?}
While no exoplanets have been detected in radio emission to date, recent
results from the study of ultracool dwarfs bode well for
future success in this area. Hallinan suggested hot Jupiters
as the most promising candidates (e.g., Hallinan et al. 2013), and that searches at
frequencies of a few 10s of MHz are likely to be the most fruitful.
These bands will be accessible with new facilities including the Long
Wavelength Array (LWA), the
Low-Frequency Array for Radio Astronomy (LOFAR), and
the Murchison Widefield Array (MWA).
Related studies will also be possible using 
a new low-frequency array under construction in Owens Valley
(with Hallinan as Principal Investigator) that will 
image the entire sky every second and reach sensitivities of
$\sim$10~mJy hourly. Completion is expected in
March 2013.

\subsection{Radio Emission from Flare Stars}
Historically, bandwidth limitations have largely precluded the study of coherent
bursts/ plasma emission on stars other than the Sun. However, this is rapidly
changing, and some spectacular new examples of stellar dynamic spectroscopy 
were presented at the meeting. T. Bastian (National Radio Astronomy
Observatory) described Arecibo
observations of a burst from the active flare star 
AD~Leo that was found to be nearly 100\%
right circularly polarized and whose emission mechanism appears to be
the cyclotron maser instability. 
The resulting dynamic spectrum showed a spectacular range of
structure, including fast-drift striae that appear to be analogous to
the auroral kilometric radiation in the terrestrial magnetosphere
(Osten \& Bastian 2008). 

G. Hallinan reported JVLA observations of the canonical flare star
UV~Ceti (a binary comprising two M5.5 dwarfs, each with a mass of
$\sim$0.1$M_{\odot}$), spanning the entire spectral range from
1-40~GHz. Both components of the binary were detected in all
bands, and one surprise is that the emission is brighter at higher
frequencies (20-40~GHz) than at lower frequencies, possibly implying
the detection of gyroresonance emission from the thermal corona. A second
intriguing finding 
was the serendipitous detection of a sweeping flare during the
observations. This flare was bright enough to detect on a single
baseline (implying the transport of significant amounts of plasma
within the corona of the star) with 100\% circular polarization. 
The flare sweeps in the opposite direction expected for a
coronal mass ejection, suggesting that it is instead a 
downward moving type~III-like burst---the first ever detected on a
star other than the Sun.

Hallinan's recent 
discoveries motivated the creation of the ``Starburst'' program at the Owens
Valley Solar Array. Led by Hallinan, it will perform night time dynamic spectroscopy
of active
stars over 1-6~GHz on a single baseline between two 27m dishes, with simultaneous optical
monitoring. Triggered follow-up with the Very Long
Baseline Array (VLBA) could then be
used to check for resolved gyrosynchrotron emission (i.e., effectively
imaging coronal mass ejections on other stars).

R. Osten (Space Telescope Science Institute) 
pointed out a puzzle concerning stellar transients
detected by {\it Swift} in hard X-rays. She noted the
case of the flare star EV~Lac, where the X-ray luminosity exceeded the
bolometric luminosity during the flare (Osten et al. 2010), suggesting that Jy-level
gyrosynchrotron flares should occur according to the G\"udel-Benz
relation. However, there is no clear
evidence yet that the radio counterpart to these types of flares has
ever been seen.

\subsection{Magnetic Chemically Peculiar Stars}
S. White (Air Force Research Laboratory) 
drew attention to magnetic CP stars as yet another category of stars
exhibiting 
electron cyclotron maser emission. He cited a recent study of CU~Vir by Trigilio et
al. (2011) that showed evidence for beaming orthogonal to the magnetic
field lines---quite different from the familiar case of Jupiter.

\subsection{The Time Domain in Cool Star Astrophysics\protect\label{time}}
The importance of the time domain in stellar radio studies was
a recurrent theme in talks spanning every topic from ultracool dwarfs to
evolved stars to novae and supernovae. 
This topic was reviewed
for the case of cool, magnetically active stars by R. Osten.

One point emphasized by Osten is that
stellar variability at mm wavelengths has yet to be thoroughly investigated.
While mm emission from young stars is typically ascribed to dust
emission from disks, some YSOs have shown spectacular mm flares that
have periodicity comparable to the orbital period. These can be
interpreted as synchrotron emission from interacting magnetospheres
based on their spectra and timescales. Such flares provide access to
the highest energy electrons from flaring stars and may impact
spectral energy distribution (SED)
fitting for YSOs. Osten reminded us that a decade ago,
Bower et al. (2003) predicted that a single exposure of Orion with
the Atacama Millimeter/Submillimeter Array (ALMA) 
that reaches $\sim10\mu$Jy sensitivity would uncover hundreds if
not thousands of flares. She noted that it will be particularly interesting to test
this prediction in light of the relatively radio-quiet behavior of
Orion stars at centimeter wavelengths described by Wolk (\S\ref{ysos1}).

Osten also drew attention to the work of
Richards et al. (2003), who found
cyclic radio flaring from the active binary HR~1099 and 
elevated states that can persist for many times longer than the
rotation period. A decade later, this phenomenon is still not
understood. This is also an example of the importance of long-term
monitoring campaigns with sampling on a range of timescales for
uncovering new phenomena.

The G\"udel-Benz relation (see above) is commonly interpreted as indicating that
there is a common reservoir out of which both particle acceleration
and plasma heating are occurring. Osten explored the question of
what happens if one also considers the time axis of this relation.
Based on data for five stars, she found that the range of
variation in radio luminosity was much larger than that of X-ray
luminosity over time, a fact that may be related to the finding that X-ray
emission appears to be saturated for ultracool stars while radio
emission does not (see also \S\ref{ultracool}).

Also relevant to time domain studies of low-mass stars,
J. Posson-Brown (Harvard-Smithsonian Center for Astrophysics)
presented a poster advertising Project Tanagra, a spectro-temporal
database of {\it Chandra} X-ray grating observations of bright
low-mass coronal stars.
These data have excellent spectral and timing resolution,
and this database is expected to provide a useful complement to radio studies of
variability in these objects.

\section{The Sun as a Radio Star}
S. White  provided an overview of radio
emission from our nearest star.
He began by reviewing the four main emission mechanisms responsible for the
production of solar radio emission: 
{\it bremsstrahlung} (thermal plasma; dominant in most of the corona);
{\it gyroresonance} (arising from strong magnetic fields in the corona, and
producing the optically thick emission observed from active regions);
{\it gyrosynchrotron} (from non-thermal electrons in flares); and {\it
  plasma
emission} (which produces highly polarized bursts at low frequency).
Electron cyclotron maser emission (discussed extensively at the
workshop in the context of ultracool stars), can be considered as a
variation of gyroresonance. As one moves across the radio spectrum,
solar radio emission goes from being optically thick coronal emission
at frequencies below 300~MHz to optically thin emission at cm
wavelengths to optically thick chromospheric emission in the mm
regime.

White noted that one of the most powerful aspects of solar radio
emission is its ability to see everything in the
solar atmosphere that can be seen at other wavelengths---and more. 
In his review talk, White also ``laid down the law'' with regard to using
the Sun as an analogy for understanding stellar radio emission,
describing how solar behavior informs what we see at radio wavelengths
from other stars. He also highlighted cases where this analogy breaks
down, as well as some of the key outstanding
problems in solar radio astronomy (see below).

\subsection{The Sun at Low Frequencies}
D. Oberoi (National Centre for Radio Astrophysics) reminded us why solar imaging is such
a challenge: a large angular size and complex morphology, a
large dynamic range in the brightness of features, and 
temporal variations on timescales ranging from milliseconds to years. 
Thus to effectively image the Sun at radio wavelengths, an
interferometric array must simultaneously deliver
high fidelity and high dynamic range imaging over a broad observing band
with high time and frequency resolution. A dense
sampling of the $u$-$v$ plane is particularly important, since time and frequency
synthesis can obscure valuable information for a dynamic source like the
Sun. Both short and long spacings are also important to resolve both the
solar disk and fine-scale structures, respectively.
Oberoi described how such performance requirements
are being met for meter wavelength science (80-300~MHz) by the MWA, 
which is currently nearing completion in the extremely radio quiet
Western Australian Outback. 

\begin{figure*}
\centering
\scalebox{0.9}{\rotatebox{0}{\includegraphics{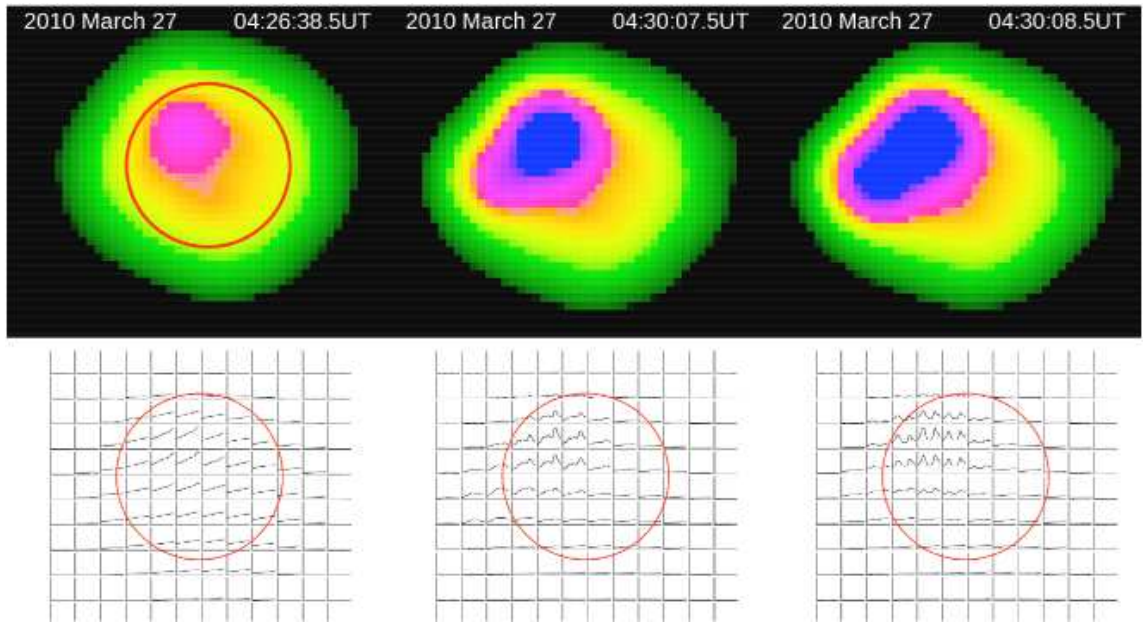}}}
\caption{{\it Top row:} Single polarization 
images of the Sun at 193.3~MHz  obtained with the MWA 32-tile
  prototype array on 2010 March 27. The data were averaged over a time
  of 1 second and a bandwidth of 0.8~MHz. The restoring beam was circular
  with a FWHM of $800''$. Dynamic range is $\sim$2500.
The red circle corresponds to the optical
  disk of the Sun. {\it Bottom row:} Spatially
  localized spectra across the disk of the Sun at times
  corresponding to the images in the top row. The bandwidth of each
  spectrum is 30~MHz. Spectra are shown for every third pixel, where
  each pixel is 
  $100''\times100''$. From Oberoi et al. 2011. }
\label{fig:MWAsun}
\end{figure*}

Oberoi presented several
examples of early science results from MWA prototype
arrays that have recently operated on site. He showed images
obtained at 193~MHz from a 32-element
engineering prototype array that exceed by an order of magnitude the dynamic
range previously achieved at similar frequencies. These data have
permitted some of the first
simultaneous imaging of thermal and non-thermal coronal emission
at meter wavelengths 
(Oberoi et al. 2011; Figure~\ref{fig:MWAsun}). Corresponding dynamic 
spectra also reveal a plethora of fine structures and
other non-thermal features not previously detectable with
existing solar spectrometers, as well as dramatic changes in the solar
plasma on timescales of less than one second. These data bode well for
the new insights into the solar corona expected
to come from the full MWA beginning in 2013. LOFAR will
complement the MWA by offering higher angular resolution observations
of the Sun at meter wavelengths, but with lower dynamic range and imaging fidelity owning to
its sparser $u$-$v$ coverage.

\subsection{The Sun at Centimeter Wavelengths\protect\label{vlaSun}}
As reported by T. Bastian,
significant progress has been made in commissioning the JVLA for solar
observing, with the first solar science observations undertaken in
late 2011. He is part of a team that has obtained the first ever dynamic imaging
spectroscopy of a decimetric type~III radio burst using the new system
(Chen et al. 2013). Such bursts are an
important diagnostic of impulsive magnetic energy release in the corona, but their
exploitation for this purpose has been previously hampered by the lack
of dynamic imaging capabilities. The high time and
frequency resolution (100 ms and 1~MHz, respectively)
and the broad frequency coverage (1-2~GHz) of the new JVLA
observations enabled tracking of the nonthermal
electron beam trajectories in the corona during the magnetic energy
release that occurred in the aftermath of a
flare. The new data imply that the spatial scale of the reconnection
sites are tens of km or less and that the coronal medium is fibrous.
The data further point to a highly fragmentary magnetic energy release process.

\subsection{Radio Emission and the Solar Cycle}
S. Cranmer raised the puzzle of why the
sphere-averaged mass flux measured from the Sun is so constant, given
that coronal heating varies by orders of magnitude over the solar
cycle (and from point to point on the Sun). 
Part of the answer appears to be that there is a sort of ``thermostat''
through which the coronal temperature is kept roughly constant via
the balance of heating conduction coming down from the corona, enthalpy
flux flowing up, and local radiation losses. These would then conspire
to keep the coronal temperature roughly constant even though coronal
heating injection rates are varying considerably.

S. White noted that the changes in the solar radio emission that do
occur with the solar
cycle are dominated by bremsstrahlung emission (and some contribution from
gyroresonance) and are much smaller than the variations
detectable in X-rays. He consequently advised that the radio domain is probably not the best
place to search for behavior analogous to the solar cycle in other
stars. However, for stars covered by kilogauss magnetic fields (and
where coronal temperatures reach as high as $\sim 10^{7}$~K), quiescent
emission could conceivably become dominated by optically thick
gyroresonant emission (a situation quite different from the solar
case). 

\subsection{The Failure of the Sun as a Role Model\protect\label{failure}}
S. White pointed out several additional areas in stellar astrophysics where our own Sun
provides frustratingly little guidance and insight for understanding
other stars.
For example, the solar wind is optically thin at all frequencies,
and to date, no stellar wind from an active dwarf has been detected in
the radio (but see also \S\ref{coolwinds}). 
For expected wind temperatures ($\sim 10^{4}$~K),
extraordinarily high mass loss rates would be required (and these are not
expected). This might change, however, if the wind/corona is in some
cases much hotter ($\sim 10^{7}$~K). Another ``disappointment'' is that
radio recombination lines have never been detected from the Sun. The
presumed explanation is that these lines are too pressure broadened to
see. V. Strelnitski (Maria Mitchell Observatory) 
suggested that a search for solar recombination lines
in the infrared might be more fruitful, as masing may make these transitions more
readily detectable.

Another puzzle pointed out by White is that on flare stars, the observed radio SEDs have
generally been interpreted as gyrosynchrotron in order to produce the
observed brightness temperatures. However, the flat spectral shapes are different
from those seen on the Sun, and it is difficult to find a plausible spread
in magnetic
field strengths that can match the data. 

\subsection{Future Prospects in Solar Radio Astronomy\protect\label{sunprospects}}
Prospects for new areas of discovery in solar radio
astronomy were pointed out by several meeting participants. 
S. White described future solar instrumentation including the planned
ultra-wide bandwidth Frequency Agile Solar Radiotelescope (FASR)
envisioned to operate from 50~MHz to 20~GHz. FASR has repeatedly
received strong
endorsements but remains unfunded. In the mean time, White noted that
solar astronomers are eagerly awaiting the commissioning of ALMA for
solar work (expected in late 2012) in order to provide imaging
capabilities for the study of the
solar chromosphere at mm wavelengths. 
Another area where ALMA is likely to contribute to the understanding
of solar emission processes is by shedding insight into flaring previously
observed in the 100-200~GHz range. As noted by White, the observed spectra show a rising
component beyond 100~GHz and thus cannot be fit by gyrosynchrotron
emission alone. A variety of explanations
have been considered (positrons, inverse Compton, thermal,
synchrotron), but none offers a satisfactory fit.

White pointed to spike bursts as examples of electron cyclotron maser
emission from the Sun. Their nature remains poorly understood, and
he hinted that further attention to these events might be prudent, 
given that they were responsible for a Global Positioning Satellite (GPS) outage
in 2006. 

S. Cranmer reminded us that the question of what drives the solar wind
is still controversial---i.e., we do not yet understand how mechanical
energy (convection) is transferred above the solar surface. One idea
is that open magnetic flux tubes are jostled about by convective
motions in the solar granulation, eventually dissipating in a kind of
turbulence. A second idea is that open magnetic fields are always
sitting in the vicinity of closed magnetic fields and that emerging
closed structures are  constantly interacting with the open
structures, causing magnetic reconnections, thereby transferring mass
and energy from the closed field to the open field. 

As illustrated by White, D. Oberoi, and T. Bastian, dynamic imaging
spectroscopy is opening a new frontier in solar
science, enabling  studies
of solar (and stellar) radio emission with unprecedented levels of
detail. However, creative new solutions are needed for managing such data sets,
where effectively every point in the time and frequency domains yields
a unique image of the Sun. Underscoring this, in the recent JVLA
commissioning study of a solar flare
described by Bastian (\S\ref{vlaSun}), 20,000 snapshots were produced
{\em per second}, and this capability has already doubled thanks to a
factor of two improvement in time resolution.

White noted that type~II solar bursts have
long been associated with coronal mass ejections (CMEs), although in his
view, this
association should be viewed with some skepticism, as examples of
type~II bursts have been seen without any obvious connection to a
CME. R. Osten countered that several studies have supported
such a link. It will be interesting to see if type~II analogs are detected
from other stars using the new generation of low frequency arrays
(LWA, MWA, LOFAR), particularly if the correlation with CMEs can be
verified. 
As pointed out by Osten, this could provide one 
means of directly observing mass loss from cool dwarfs, whose
``steady'' winds are generally too feeble to be directly detected 
($\sim 10^{-14}~M_{\odot}$ yr$^{-1}$).
While CMEs are only a minor contributor to mass loss from the Sun,
they may be more important for other types of stars, particularly
active young stars.

Solar noise storms remain an enigma, in part
because no counterparts have been identified at other
wavelengths.  They are known to sit over vigorous active
regions, but how they manage to produce a steady flux of energetic electrons is unknown.
Analogs to noise storms have also now been identified on other
stars (e.g., RS~CVn). 
White showed a dynamic spectrum (52-71~MHz) of a solar noise storm
recently obtained with the LWA.
However, noise storms are more often seen at higher frequencies, and a recent
example detected from 1-2~GHz with the JVLA by Chen \& Bastian (in
preparation) was also presented. 

\section{Stellar Winds and Mass Loss}
Virtually all stars are believed to be losing mass through stellar
winds, although the rate of this mass loss varies by nearly 12
orders of magnitude across the H-R diagram. This mass loss impacts the
evolution of the stars themselves, the star's surroundings, and
ultimately, affects the evolution of the entire Galaxy. As we were
reminded in the review by S. Cranmer, the key
driving 
mechanisms for stellar winds include: (1) gas pressure; (2) radiation
pressure; (3) wave pressure/shocks; (4) MHD effects (which may produce
mass loss through CMEs or ``plasmoids''). Molecular line emission
(from cool giants) and
free-free emission (from partially ionized winds of hot stars) are two ways in which
radio observations have traditionally been used to study stellar mass
loss. 

\subsection{Hot Star Winds}
We were reminded by  M. G\"udel and S. 
Kwok that hot star winds are a case where radio observations
led to a whole new concept in stellar mass loss. The winds from these
stars are ionized and give rise to optically thick thermal emission in
the radio. By observing at shorter and shorter radio wavelengths, one
sees material closer and closer to the star. Classically, the flux density from
these stars is predicted to scale as $S_{\nu}\propto({\dot
  M})^{\frac{4}{3}}\nu^{0.6}$ where ${\dot M}$ is the mass loss rate
of $\nu$ is the frequency. However, mass loss rates derived this
way are systematically higher than those derived from other
diagnostics (e.g., X-ray line asymmetries), implying that the wind is
likely clumped, making its emission proportional to the square of the
density.

Related to this theme, a poster by M. M. Rubio-D\'\i ez (Centro Astrobiologia, INTA-CSIC)
focused on the reconciliation between the mass loss rates
of OB star winds derived from a variety of tracers. Correctly
accounting for density inhomogeneities is found to be crucial, and she
emphasized the importance of sampling the entire wind 
using a range of tracers (including radio
observations of the outermost wind region) for constraining models.

The current status of stellar wind theory was reviewed by
S. Cranmer, who described a number of aspects of stellar wind
physics that are important to keep in mind when
interpreting observations. For example, 
he highlighted the effect of ``gravity darkening'' in fully
radiative stars. Gravitational effects alone would predict more mass
loss from the equatorial regions of stars, where centrifugal forces are
maximized; however, gravity darkening actually reverses this trend
(because the local effective temperature is much higher at the poles
than at the equator),
leading to enhanced polar mass loss and to the production of
prolate, bipolar outflows. While gravity darkening likely cannot
explain all such outflows, Cranmer suggested that it may be an important and overlooked
effect in some instances.

Cranmer reminded us that stellar winds play a fundamental role in massive star
evolution and that the fraction of the total mass
returned to the ISM through winds vs. supernova explosions changes
dramatically as a function of initial stellar mass. For $\sim$8-10~$M_{\odot}$ stars, most
of the mass return occurs in the subsequent supernova explosion; mass loss during
the red supergiant phase becomes increasingly important for
$\sim$20-30~$M_{\odot}$ stars, while for $M_{\star}\gsim30M_{\odot}$, the
Wolf-Rayet phase becomes  dominant in the mass recycling.

Several puzzles persist concerning mass loss and winds from classical Be
stars, including the formation mechanism for their ``decretion'' disks. 
Since these stars are rotating well below breakup speed, it
remains unclear  how their circumstellar gas acquires angular
momentum, although non-radial pulsations are likely to be involved in
providing the spin-up needed to effectively cancel gravity
(e.g., Cranmer 2009).  Cranmer noted that Be stars also lose mass through ionized polar
winds, but the ionization source for these outflows is
unknown. Unstable shocks are one possibility.

\subsection{Winds from Cool Dwarfs\protect\label{coolwinds}}
Constraining the driver of mass loss from cool dwarfs ($T_{\rm
  eff}\lsim$8000~K) has been notoriously difficult since the feeble
winds from these stars tend to be too low to detect directly (see also
\S\ref{failure}, \S\ref{sunprospects}). This makes
insights gleaned from the solar wind particularly important.
Cranmer described recent theoretical work (Cranmer \& Saar 2011)
that for the first time provides physically motivated predictions 
for the stellar winds of 
late-type stars that enable prediction of mass loss rate directly
from a star's fundamental properties (including rotational velocity). 
This work takes inspiration from recent progress
in understanding solar wind acceleration by tracing the energy flux
from MHD turbulence from a subsurface convective zone to its
dissipation and escape through open magnetic flux tubes as a stellar wind. 
Results are a significant improvement over past semi-empirical
correlations which could not reproduce star-to-star variations caused
by differing rotation rates.

Cranmer stated that while stellar wind theory is typically successful
to within an order of magnitude, one must keep in mind that applying
these theories requires knowing considerable information about the star
in question (e.g., mass, luminosity, age, rotational velocity,
magnetic field strength, pulsation properties). Fortunately, as we
heard at the meeting, radio observations of stars offer one means to
constrain many of these parameters. 

Finally, Cranmer raised the questions of whether ``multi-thermal'' stellar winds
might exist, analogous to the multi-phase structure of the interstellar
medium (ISM). Answers could come from, e.g., combining radio (and IR)
observations of ``cold'' dust-forming gas  with UV spectroscopy to
probe warmer chromospheric gas.

\subsection{Winds from Red Giants}
E. O'Gorman (Trinity College Dublin) 
reminded us that despite decades of research, the driving
mechanism of the winds from ``ordinary'' (non-pulsating) Red Giants remains a puzzle. 
Lying beyond the Linsky-Haisch dividing line (Linsky \& Haisch
1979), these stars have
minimal chromospheric emission, but instead exhibit slow, relatively dense,
largely neutral winds where $V_{\rm terminal}<V_{\rm escape}$. This implies
that most of the energy of the wind goes into lifting it out of the
stellar gravitational potential, not the final wind kinetic energy.
Since these winds have insufficient dust and molecular opacity for
efficient radiative driving, and since mass-loss rates too high to be
explained by pulsationally-driven wind models, and since hot
wind plasma is absent (ruling out ``coronal'' type winds), some type of
magnetically-driven wind is favored. However, many questions remain.

One key to constraining the wind driving mechanism is accurate
characterization of the temperature and density structure of the
wind. While these are poorly constrained by traditional ultraviolet
observations, radio observations in the mm/cm can provide key
insights, since the thermal continuum emission depends linearly on
temperature. Another advantage of radio observations is that the
opacity scales as roughly $\lambda^{2.1}$; consequently, longer
wavelengths probe further out in the wind, effectively allowing
multi-frequency observations to provide spatial information on objects
that are unresolved. The drawback has been that Red Giants
(whose centimetric emission stems from thermal free-free from the
ionized component of the wind) 
are feeble radio emitters, making them too weak to detect at a wide
range of radio frequencies---until now.

Thanks to the greatly expanded bandwidth of the JVLA, O'Gorman
showed that it is now
possible to obtain SEDs for nearby Red Giants over the full range of
JVLA bands from 3 to 43~GHz. He presented results for two stars, $\alpha$~Boo
(Arcturus) and $\alpha$~Tau (Aldebaran), and for the first time, complete
frequency coverage was obtained for each star within a period spanning
less than two weeks, thus mitigating effects of possible variability. 
Both stars were detected as point sources in all bands with flux
densities ranging from 0.06 to 3.7~mJy, and the new measurements reveal
discrepancies with existing models. To investigate the cause of this,
O'Gorman plans to employ future
modeling using a new hydrogen ionization code (which accurately treats
the ``freezing out'' of the ionization balance in the wind) to
provide new insights into temperature and density profiles of the
winds as a function of
radius.

\section{Evolved Stars}
Radio observations have long been recognized as a powerful tool for
studying atmospheric physics, mass loss, and circumstellar chemistry
during the late stages of stellar evolution. Not surprisingly, an entire day of the
Radio Stars workshop was devoted to discussion of the latest developments
in the study of red supergiants (RSGs), 
asymptotic giant branch
(AGB) stars, and the evolution of the latter into
planetary nebulae (PNe).  Cepheid variables were also briefly touched
upon. An overview talk by R. Sahai (Jet Propulsion
Laboratory) set the stage
for the discussion of this broad topic (see below).

\subsection{Red Supergiants\protect\label{supergiants}}
Observations were presented at the workshop for a diverse set of red
supergiants (RSGs; see also \S\ref{photospheres} and
\S\ref{linesurveys}). 
B. Zhang (Max Planck Institut f\"ur Radioastronomie) 
described recent VLBA and VLA observations of NML~Cyg, one of
the most luminous RSGs in the Galaxy. Based on astrometric
measurements of the H$_{2}$O masers over several epochs, he and his
colleagues obtained a distance and proper motion consistent with this
star being a member of the Cyg~OB2 association (Zhang et al. 2012).
Using the VLA, they detected the radio photosphere of the star at
43~GHz and measured a diameter of 44$\pm$16~mas. 
This result would imply an enormous size for the star
($\sim$70~AU for a distance of 1.6~kpc). Future JVLA measurements
will be able to confirm this result and to significantly reduce the error bar on this
measurement. A combination of the VLBA and VLA data also allowed accurate
astrometric registration of the SiO and H$_{2}$O masers with the
photosphere of the star for the first time, revealing a highly
asymmetric distribution of masers that is inconsistent with past
models of the source geometry.

A. Richards (Jodrell Bank) presented new 5.75~GHz imaging observations of Betelgeuse
($\alpha$~Orionis) obtained with the expanded Multi-Element Radio
Interferometry Network (e-MERLIN) in 2012 July
with $\sim$180~mas resolution. While analysis was still ongoing as of the
time of the meeting, initial results showed the presence of two
emission peaks within the central image of the star, as well as
emission ``hot spots'' with brightness temperatures in excess of
4000~K. Additionally, the data revealed evidence for a ``plume'' extending
to the southwest, similar to what has been seen independently by other
workers in
optical interferometric imaging and CO
imaging at mm wavelengths.

V838~Mon, famous for its light echo, 
erupted in a nova-like burst in 2002 and emerged as a
10$^{6}~L_{\odot}$ RSG star 
six months later. The star is believed to have formed as 
the product of a recent merger of two main
sequence stars. How long its supergiant phase will last is a 
question being studied by S. Deguchi (Nobeyama Radio
Observatory). Deguchi  and his
colleagues monitored the star for SiO masers from Nobeyama, first detecting the SiO
$v$=1 and $v$=2, $J$=1-0  lines roughly 3 years after the 2002 outburst
(Deguchi et al. 2005). Since then they have
found the intensities of the lines to be decreasing
gradually. 
Deguchi emphasized that continued monitoring of the SiO masers is important
for providing insight into what fraction of SiO maser stars (and what
fraction of RSGs)  are likely to be merger products.

While most RSGs are field stars,
Deguchi also described observations of SiO masers from RSGs in massive young star
clusters, which can offer important laboratories for understanding RSG
evolution. He and his colleagues have been using
such observations to derive velocity dispersions
for the clusters, as well as dynamical masses and kinematic distances
(e.g., Fok et al. 2012). One surprising finding was that the masses of RSGs in
massive clusters are found to be smaller on average than those of some
well-studied RSGs such as VY~CMa. 

\subsection{Radio Photospheres\protect\label{photospheres}}
K. Menten (Max Planck Institut f\"ur Radioastronomie) 
highlighted the promise of the JVLA and ALMA for the study
of radio photospheres of red giants. He cited as examples the possibility to obtain
more than an order of magnitude improvement in accuracy in the
photospheric diameter of IRC+10216 compared with recent VLA
measurements (Menten et al. 2012) and the ability to image star spots (if they exist) in
the radio photosphere of Betelgeuse (see also \S\ref{supergiants}). 
In the case of ALMA, he noted
that study of the composition
of the molecular photosphere will become feasible (see
\S\ref{chemistry}) and that 
the $\sim$20\% size variations in AGB stars due to pulsation
are expected to be discernible. 
M. Reid (Harvard-Smithsonian Center for Astrophysics) also reminded us that
existing observations of radio photospheres place stringent limits on
the presence of regions of hot ($\sim$10000~K)
gas at a few stellar radii in AGB stars, seemingly ruling out strong
shocks in their extended atmospheres.

\subsection{Mass Loss on the Asymptotic Giant Branch}
\subsection{Mass Loss Rates and Geometries\protect\label{massloss}}
As pointed out by R. Sahai, a 
limitation of past studies that have derived mass loss rates for
AGB
stars using molecular line observations (primarily CO) is that they
have relied on simple or analytical models in which some kinematic
temperature was assumed. Sahai highlighted as a 
major advance in the past few years the
emergence of new studies that interpret molecular line data
using full self-consistent radiative transfer and thermodynamic
modeling (e.g., De Beck et al. 2010).

\begin{figure}
\centering
\scalebox{0.6}{\rotatebox{0}{\includegraphics{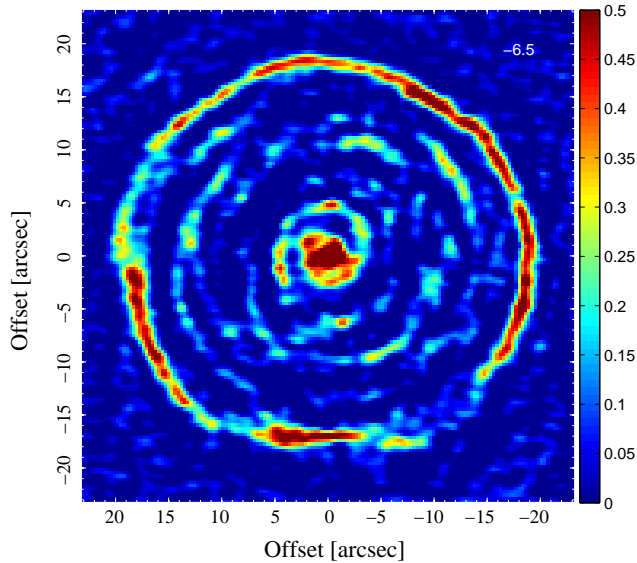}}}
\caption{CO(3-2) emission at the stellar systemic velocity ($V_{\rm
    LSR}=-6.5$~\kms)  from the
  circumstellar environment of the AGB star R~Sculptoris, observed
  with ALMA during Cycle~0 (Maercker et al. 2012). 
The intensity scale is in Jy beam$^{-1}$. }
\label{fig:deathspiral}
\end{figure}

While the assumption of smooth, spherically symmetric outflows from
AGB stars works reasonably well in many cases, it is also seen to
break down in some instances. One spectacular
example was presented by M. Maercker (European Southern Observatory),
who showed data for the carbon star R~Scl recently obtained with 
ALMA (Maercker et al. 2012). R~Scl was known
from previous {\it Hubble Space Telescope} imaging to be surrounded by a detached shell,
thought to be the product of a thermal pulse. However, recent ALMA
CO(3-2) observations
revealed a surprise: a 3-D spiral pattern extending from the star to the
detached shell (which is itself also detected in CO;
Figure~\ref{fig:deathspiral}). 
The spiral pattern betrays the
presence of an unseen companion that is sculpting the
outflowing AGB wind. Imprinted in this spiral
pattern is a historical record of the stellar mass loss history since
the last thermal pulse. 

Maercker showed that from
simple geometric arguments alone it is possible to derive the orbital period of the
binary, the wind expansion velocity as a function of time, the
evolution of the mass loss rate with time, and the age of the detached
shell. 
The spherically symmetric shape of the detached shell also
constrains the pulse duration to be much shorter than the binary
orbital period. One curiosity was that no evidence for
deceleration of the shell was seen, as would be expected as a
consequence of interaction
with debris from previous mass loss. Maercker's data set
represents the first {\em observational} constraints on the pre- and
post-thermal pulse state of an AGB star; until now, our
existing understanding of thermal pulses has been essentially
theoretically driven. This is significant since nucleosynthetic
yields from AGB stars depend critically on the duration and frequency
of thermal pulses. 

S. Kwok showed another example of distinctly non-spherical
structures in the circumstellar envelope (CSE) of the carbon star CIT~6. Employing
spatiokinematic modeling, a recent study by Kwok and his
colleagues (Chau et al. 2012) unveiled the presence of several 
incomplete shells. Similar features have also been previously 
seen in IRC+10216. These partial
shells may be due to time- and space-dependent mass loss, but the
underlying driving mechanism remains unclear.

\subsubsection{Probing the Complete History of AGB Mass Loss}
R. Sahai highlighted the challenge of determining the total mass of
CSEs of evolved stars given that commonly used molecular diagnostics constitute only trace
components of the CSE and 
do not sample the entire envelope. One means
of addressing this question is through observations of circumstellar
atomic hydrogen.
T. Le~Bertre (Observatoire de Paris) described the range of advantages the 
\HI\ 21-cm line offers for studying the extended portions of CSEs 
and hence the extended mass loss history of AGB and related
stars. Because \HI\ is not readily destroyed by the interstellar
radiation field, it can be used to probe the circumstellar ejecta to much larger
distances from the star compared with commonly used molecular tracers
such as CO, thereby tracing a larger fraction of the stellar mass
loss history ($>10^{5}$~yr). As a spectral line, it also provides valuable kinematic
information. Moreover, because in most instances the hydrogen is
optically thin ($h\nu/kT<<1$), this implies that \HI\ measurements
can be directly
translated into a measurement of the mass in atomic hydrogen (assuming
the distance is known).

Le~Bertre described an ongoing program at the Nan\c{c}ay Radio
Telescope (NRT) that has now surveyed more than one hundred AGB and related
stars in the \HI\ 21-cm line (G\'erard et al. 2011). The target stars span a range in
chemistries, mass loss rates, variability classes, and other
properties. Detection rates are high ($\sim$80\%) for semi-regular and irregular
variables, but lower for Miras despite their on-average higher mass loss
rates. This paradox may be due to the generally lower
temperatures of the Miras, which may lead to more hydrogen in molecular
form or atomic hydrogen that is too cold to be readily detected in emission. 

One discovery from the NRT studies is that
the observed \HI\ line profiles are frequently narrower than seen in
CO and in some cases have a two-component structure that includes a broader
pedestal. This finding can be explained by a freely expanding wind
that is slowed down at large radii by the surrounding medium (e.g., Le~Bertre et
al. 2012). The narrow line
component then results from a highly extended ($\sim$0.1 to 1.0~pc), 
quasi-stationary shell of
material  that accumulates between a
termination shock and the ISM interface, while the broader component is the
counterpart to the freely expanding wind at smaller radii. Another
finding is that the \HI\
line centroids of the detected stars are frequently found to be
displaced toward zero Local Standard of Rest velocity,
indicative of interaction between the CSE and the local ISM owing to
the star's space motion.

\begin{figure}
\centering
\scalebox{0.47}{\rotatebox{0}{\includegraphics{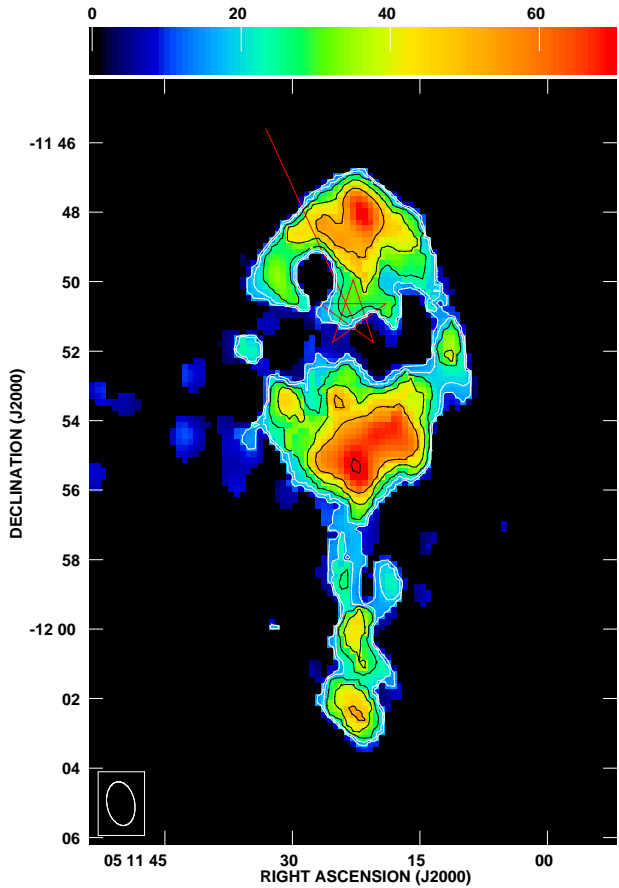}}}
\caption{\HI\ total intensity image of the AGB star RX~Lep from Matthews
  et al. 2013. The star symbol indicates the stellar position, and the
  red line indicates the direction of space motion. The projected
  extent of the
  \HI\ shell and tail is $\sim$0.7~pc in the north-south direction.
The intensity scale has units of Jy beam$^{-1}$ m s$^{-1}$. The
    synthesized beam is $\sim 76''\times 48''$.  }
\label{fig:rxlep}
\end{figure}

Following Le~Bertre's talk, I described an ongoing program with the
VLA and JVLA to image circumstellar \HI. 
To date, we have successfully imaged more than a dozen red giants 
(e.g., Matthews et al. 2013). In all cases we find clear
evidence that the properties of the extended CSE are set by both the
properties of the star itself and by the interaction between the star and
its interstellar environment.  One 
manifestation of this is the discovery of several examples of \HI\ ``tails'' up to a
parsec or more in extent, trailing the
motions of the stars through space  (Figure~\ref{fig:rxlep}). These tails are
predominantly seen associated with stars with high space velocities ($V_{\rm
  space}\gsim$60~\kms) and arise from
ram pressure on the circumstellar ejecta as these mass-losing stars
move supersonically through the ISM. 
In some instances, we have detected velocity
gradients along these \HI\ wakes that arise from deceleration of the
gas through interaction with the ISM; 
such measurements in turn supply a means of age-dating the mass
loss history of the star, confirming that the \HI\ traces $>10^{5}$
years of stellar mass loss (e.g., Matthews et
al. 2011). For stars with lower space velocities, interaction with the
local ISM is also seen manifested through the presence of 
detached \HI\ shells, formed when the outflowing wind is abruptly
slowed at a termination shock. The presence of these shells in our
images confirms
the picture first inferred from the two-component NRT line 
profiles described by Le~Bertre. One example is our recent
discovery of an \HI\ shell with a diameter of $\sim$0.24~pc 
surrounding Betelgeuse (Le~Bertre et al. 2012). This
result was showcased  in a poster presented 
by Le~Bertre.

\subsection{Mass Loss from Cepheid Variables\protect\label{cepheids}}
Despite the importance of Cepheid variables as extragalactic distance
indicators and as cornerstones to our understanding of the evolution of
intermediate mass stars,
a puzzling ``Cepheid mass discrepancy'' has persisted for
several decades: the masses inferred from stellar evolution models are
systematically higher than those derived from stellar pulsation (or
binary orbital dynamics when available). Mass loss has long been proposed 
as a solution to this discrepancy, but direct empirical confirmation has
remained elusive. I described how new insights are coming from
observations of Cepheids in 
the \HI\ 21-cm line. Using the VLA, we recently detected an extended
($\sim$1~pc across)
\HI\ nebula surrounding the position of the Cepheid archetype,
$\delta$~Cephei (Matthews et al. 2012). The position and head-tail morphology of this
emission are consistent with a remnant of recent/ongoing mass loss from
$\delta$~Cephei. Based on these data, we have derived a mass loss rate
${\dot M}\approx(1.0\pm0.5)\times10^{-6}M_{\odot}$ yr$^{-1}$ and an outflow
velocity of $V_{\rm out}\approx$35~\kms\ (the first ever measured for
a Cepheid). Analysis of data from a JVLA survey of additional Cepheids in the \HI\
line is ongoing.

\subsection{Circumstellar Masers\protect\label{masers}}
\subsubsection{Observations}
S. Deguchi reviewed the topic of circumstellar masers from an observational
perspective. He reminded us that in addition to the maser species
commonly observed in the atmospheres of oxygen-rich evolved stars 
(OH, H$_{2}$O, SiO), masers are also seen in carbon-rich
stars, including HCN, SiS, and CS. The latter are typically weaker
than the masers in oxygen-rich stars, and consequently have been
relatively little studied. He also underscored that stellar masers are
not just limited to AGB stars, but also occur in a variety of more
``extreme'' stellar environments where they can provide important new insights
into sources ranging 
from RSGs (\S\ref{supergiants}) to
water fountains (\S\ref{fountains}) to novae (\S\ref{novae}).

Two poster contributions presented new results on the molecular masers
in Mira ($o$~Ceti). G. McIntosh (University of Minnesota, Morris)
presented 
long-term monitoring data from the Haystack and Mopra telescopes of
multiple SiO transitions. Evidence was seen for correlations between
the velocity range of emission and stellar phase. R. Zizza (Wellesley
College) presented a complementary study that showcased six
epochs of simultaneous VLA imaging data of the SiO $v$=1, $J$=1-0 and
22~GHz H$_{2}$O masers in Mira. Consistent with McIntosh's results, the integrated
flux density of both the SiO and H$_{2}$O masers is seen to correlate
with the stellar phase-dependent variations in the optical light. 
These data also represent some of the first ever concurrent
imaging of the SiO and H$_{2}$O masers in an AGB star, providing
direct information on their relative locations in the stellar envelope.

Interferometric imaging of the water masers in the atmospheres of AGB
stars (including Zizza's VLA study of Mira) have shown that these
masers tend to lie in ``shells'' with radii $r\sim$5-30$R_{\star}$. 
A. Richards showed that this
relationship also holds for RSG stars despite the fact that the stellar radius
grows from $\sim$1~AU for AGB stars to roughly an order of magnitude larger for
RSGs.

Theoretically, brighter water maser clumps in evolved stars are expected to exhibit
smaller beamed sizes (assuming they are spherical). However, Richards
described recent work based on MERLIN observations in which she and her colleagues 
identified two stars in which this prediction is
violated. The two stars exhibit the most extreme water
maser variability in their sample, suggesting that masers arise instead from shocked slabs
(Richards et al. 2011). Their recent work also shows that in RSGs, 
water maser clouds can be traced for upwards of 5 years,
while in AGB stars the lifetimes tend to be less than 2 years. These
survival times are comparable to sound crossing timescales but
much less than the maser shell crossing times. While results from a
combination of interferometric and single-dish monitoring do show
velocity  peaks in the spectra vanishing and re-appearing, Richards
proposed that this was due to the masers turning off from turbulence or
beaming effects, not the actual destruction and reformation of clouds,
since there is not an obvious mechanism by which this could occur.

Richards also presented observations of several stars where
numerous individual water maser clouds are seen to
brighten/dim in lockstep when observed at subsequent epochs. Such
correlations visible on the timescale of weeks seems to rule out
shocks as responsible for the fluctuations, since the shock
propagation velocities required would have to be unphysically
high. Instead this observation seems to point toward a radiatively
driven pumping mechanism.
Finally, Richards is finding evidence that water maser cloud size, $R_{c}$, is
proportional to the stellar radius
[$R_{c}\sim(0.7\pm0.3)R_{\star}^{1.0\pm0.1}$]. One interpretation is
that convection cells determine the cloud size. 

\subsubsection{Maser Theory}
V. Strelnitski 
noted that the fundamentals of the theory of maser
amplification in astrophysical sources were developed in the late 1960s
and early 1970s (soon after the discovery of cosmic masers) and that
advances in maser theory have since somewhat stagnated. His
review talk nonetheless pointed out some new developments and also highlighted several
concepts that are crucial for astronomers to be mindful of 
as they attempt to gain physical insights from maser observations.

One point emphasized by Strelnitski is that a maser requires a system
with more than two population levels (a two-level system cannot
sustain the population inversion required to produce a maser since its
excitation temperature always remains positive). This has consequences
for assessing  when a maser ``thermalizes'', as one
cannot simply compute the thermalization condition for a maser by using a
two-level approximation; doing so may produce large
errors in the maximum allowed gas densities where a maser may
occur. For example, if collisions participate in pumping,
thermalization will occur
at much higher densities than in a two-level system.

Concerning the characterization of maser pumping mechanisms,
Strelnitski stressed the need to consider not just the ``source''
(i.e., a
high-energy reservoir from which energy is extracted), but 
both the {\em source}  and the
{\em sink} (the low-energy reservoirs to which energy is
transferred). Thus maser pumping is never just ``collisional'', but e.g.,
``collisional-collisional'' or ``collisional-radiative''. He cited the
latter as being the most likely to occur in CSEs, with collisions
pumping the molecules to higher levels and spontaneous emission
creating the ``sink'' that completes the cycle. 

Strelnitski cited  several areas where additional work would be
valuable. One example is extending the use of masers as tracers
of turbulence. A promising application may be to the study of
water fountain sources (\S\ref{fountains}). He also advocated work on
the development of methods of extraction of the principal pumping
cycles in numerical calculations,  which will enable recovery of
the details of maser pumping schemes (e.g., Gray 2007). 
At present, most numerical schemes typically provide
only a final solution, and no details on the population flow are
stored.

\subsection{Magnetic Fields in Evolved Stars}
The number of AGB (and post-AGB) stars whose magnetic fields have been
characterized is small but rapidly growing.
W. Vlemmings (Onsala Space Observatory) provided an overview of this topic 
and the observational techniques being employed to
probe magnetic fields and their evolutionary roles in evolved stars. 

To date, the bulk of measurements of magnetic field {\em strength} in
evolved stars and PNe have been derived
from circular polarization (Zeeman splitting) measurements of OH,
H$_{2}$O, and SiO masers, or
in a few cases, Zeeman splitting of non-masing molecular
lines. Constraints on the {\em shape} of the magnetic field come from linear
polarization measurements of masers as well as observations of aligned dust grains and 
molecules (via the so-called Goldreich-Kylafis effect). 

As described by Vlemmings, recent measurements of SiO masers in oxygen-rich
stars (which lie at radii
$r\sim2R_{\star}$) suggest magnetic field strengths of $\sim$1-10~G,
with the caveat that there are other mechanisms that might explain the
observed polarization without requiring such a strong field. Water
masers (lying between $\sim$50-80~AU) reveal field strengths 
$B\sim$0.1-2~G, and OH masers (lying at $\sim$100-1000~AU) imply
$B\sim$1-10~mG. For carbon-rich stars, a handful of measurements of CN
at $r\sim$2500~AU yield $B\sim$7-20~mG.  The radial
dependence of the magnetic field is still not well constrained owing
to uncertainties in the location of the masers relative to the central
star, and
observations are consistent with either $r^{-2}$ (solar-type)
or $r^{-1}$ fall-offs. Despite this uncertainty, the
results to date point to magnetic
fields dominating the energy densities of AGB stars out to radii of
$\sim$50~AU.

Vlemmings reported new VLBA measurements of
H$_{2}$O masers in 3 AGB stars which point to field strengths of a few
hundred mG,
consistent with previous measurements of other AGB stars. 
They have also detected for the first time in AGB stars
weak linear
polarization in the H$_{2}$O lines (at the $\sim$1\% level; 
Leal-Ferreira et al., 2012a). 
Previously linear polarization of the H$_{2}$O masers had been
detected only in PPNe and star-forming regions.
For the proto-planetary Rotten Egg Nebula (see also \S\ref{egg}), Vlemmings and 
colleagues measured $B\sim$45~mG and derived an extrapolated magnetic field
strength of $\sim$3~G. This represents one of the first magnetic field measurements
around a known binary system (Leal-Ferreira et al. 2012b). 

Vlemmings described recent developments in the study of magnetic fields using
non-masing (thermal) lines via the Goldreich-Kylafis effect. This approach
complements maser measurements by more comprehensively probing the
entire CSE. Demonstrating the feasibility of this method, he showed
new Submillimeter Array (SMA) 
polarization measurements of several thermal lines toward AGB stars
including IRC+10216 
(Girart et al. 2012), IK~Tau (Vlemmings et al. 2012) and $\chi$~Cyg
(Tafoya et al., in preparation).  In all cases, evidence of non-radial linear 
polarization is observed, consistent with the presence of large-scale
magnetic fields.

N. Amiri (University of Colorado, Boulder) 
reported recent VLBA measurements of the SiO masers and their
polarization properties in the OH/IR star OH~44.8$-$2.3. This is the
first OH/IR star whose SiO masers have been studied with Very Long
Baseline Interferometry (VLBI)
resolution. The SiO
masers in OH~44.8$-$2.3 were found to lie in a ring pattern with a
radius comparable to those measured in Mira variables (Amiri et
al. 2012). This result is a surprise, since the central star is
expected to be much larger than an ordinary Mira. The maser ring also
showed high fractional linear polarization (up to
$\sim$100\%) and a geometry consistent with a dipolar field. Gaps in
the SiO ring as well as an elongation in the OH masers found in
archival VLA observations reinforce this interpretation and suggest
that magnetic fields may impose a preferred outflow direction in the
CSE. This deviation from spherical symmetry is of
particular significance, as
OH/IR stars are thought to be nearing the transition to the PN stage.

Despite impressive progress, Vlemmings emphasized that much work remains to be done in
characterizing magnetic fields and their role in late stellar
evolution, and measurements of larger samples of stars will be
critical for answering many of the outstanding questions. These include the
ubiquity of magnetic fields, the impact
of binarity on the magnetic field, the dynamical importance of the
field, their role in AGB mass loss, and the origin of the field itself
(e.g., a single star dynamo, a binary star or heavy planet, or disk interaction).

\subsection{The Road to Planetary Nebula Formation\protect\label{PNe}}
\subsubsection{The Origin of Non-Spherical Geometries}
Less than 5\% of planetary nebulae (PNe) appear spherical, and
the origin of the spectacular diversity of morphologies of 
PNe has remained a topic of vigorous research for decades. R. Sahai 
summarized results to date that point toward the mechanism responsible
for shaping PNe becoming operational well before the PN 
phase itself---either in the post-AGB phase or possible very late in the AGB.
S. Kwok also reminded us that one of the reasons radio observations of
PNe are particularly valuable is that the bulk of their mass lies beyond what can
be traced via optical light.

As described by Sahai, the collimated fast winds/jets that appear during the late
AGB/post-AGB are believed to be the primary agent shaping PNe,
although dusty equatorial tori may also help to confine the
flows. Sahai noted that the fast bipolar outflows of
proto-planetary nebulae (PPNe) have a large momentum excess ($L/c <<
dM/dt \times V_{\rm exp}$ ), implying the winds cannot be radiatively
driven. Here $L$ is the stellar luminosity, $c$ is the speed of light,
$dM/dt$ is the mass flux and $V_{\rm exp}$ is the expansion velocity
of the wind. The driving mechanism of these outflows thus remains
unclear. J. Sokoloski (Columbia University) also
pointed out that several symbiotic stars (which are known binaries)
exhibit morphologies similar to PNe, offering additional support for
the idea that binarity is a key shaping factor.

Sahai described a recently published survey of mass loss from a sample
of post-AGB stars and young PNe using data from Owens Valley Radio
Observatory (S\'anchez Contreras \&
Sahai 2012). In total, 24 of 27 targets were detected in at least one
CO transition, including 11 for the first time. Asymmetries and velocity
gradients were seen to be the norm, with broad, high-velocity line wings
indicative of fast outflows in more than 50\% of the sample.

D. Tafoya (Onsala Space Observatory) presented a poster describing
results from a study of
two ``water PNe'', which based on the presence of water molecules, are
believed to be in the very earliest stages of the PN phase. Such
objects therefore offer an opportunity to study the development
of bipolar PN morphologies. These objects are sufficiently young
to still possess their molecular envelopes from the AGB mass loss
phase. The source
IRAS17347-3139 was detected with the SMA, and 
the $^{12}$CO emission was found to exhibit three
components; a slow extended component ($\sim$30~\kms) and two
high-velocity compact components ($\sim$150~\kms). The slow component
is roughly spherical and likely originated during the AGB, while the
high-velocity components may be part of a precessing bipolar wind.

\subsubsection{The Origin of Dusty Equatorial Components}
R. Sahai cited  the origin of dusty equatorial components as one of the
greatest puzzles concerning post-AGB stars and PNe. This includes both ``dusty
waists'' with sizes $\sim$1000~AU seen in PPNe and $\lsim$50~AU
circumbinary disks surrounding binary post-AGB stars. Recent work (Sahai et al. 2011) 
shows sub-mm excesses in both classes of objects, implying the
unexpected presence of
large grains. The long formation timescales required for these large grains place 
constraints on the age and formation of the dusty equatorial
regions.

\subsubsection{Distance Measurements to Planetary Nebulae\protect\label{egg}}
Accurate distance determinations for PPNe are crucial for constraining
their evolution and fundamental parameters. Y. Choi
(Max Planck Institut f\"ur Radioastronomie)
described VLBA parallax and proper motion measurements of the water
masers associated with OH~231.8+4.2 (also known as the
``Rotten Egg'' nebula), a PPN with a bipolar optical morphology and a
binary central star. The resulting distance has an accuracy of
better than 2\% (1.54$\pm$0.02~kpc; Choi et al. 2012). Her team found
blueshifted and redshifted
components to the north and south, respectively, with a velocity
separation of $\sim$20~\kms. Thus the outflow traced by the H$_{2}$O
masers is consistent with an outflow from the central AGB star. This
contrasts with observations in CO and
other tracers previously measured at larger distances from the central
star, where velocity differences between the two lobes are an order of
magnitude higher.

\subsubsection{The Coldest Object in the Universe}
The PPN known as the Boomerang has been referred to as 
the ``coldest object in the Universe'' after a 
study by Sahai \& Nyman (1997) showed that the kinematic
temperature of the CSE is less than 2~K due to adiabatic expansion
and a mass-loss rate that is sufficiently high
($\sim10^{-3}~M_{\odot}$~yr$^{-1}$) to makes the CO(1-0) line optically thick
to heating by the cosmic microwave background. Sahai presented new 
ALMA CO(1-0) and (2-1) observations of this object that
reveal a bipolar outflow with the same orientation as the outflow previously
seen in the optical. Evidence for a dense central waist (with
indications of mm-sized grains) is seen along
with lobes that exhibit a bubble-like structure. 
Weak, patchy absorption from the ultracold shell 
is also visible, together with patchy emission on the periphery of the
shell. The latter provides the first direct evidence of grain photoelectric
heating in an AGB star's CSE. 

\subsubsection{Water Fountain Sources\protect\label{fountains}}
Water fountain sources are post-AGB stars (or very young PNe) with
22~GHz water maser emission spread over large velocity ranges, indicating
high-velocity outflows ($>$100~\kms). Roughly ten examples of this
class are presently known. S. Deguchi described recent VLBA
observations of the H$_{2}$O masers in one example,
IRAS18286-0959, where he and his colleagues identified two bipolar
precessing jets that form a double helix structure (Yung et al. 2011) .  This is almost
certainly the hallmark of a binary, reinforcing the long-suspected link
between water fountain sources and binarity. However, it remains unclear where the
jets are launched (e.g., in an accretion disk around a compact object,
the atmosphere of the post-AGB star, or both).

\subsection{Circumstellar Chemistry\protect\label{chemistry}}
\subsubsection{Overview}
To date, roughly 80 molecular species have been detected in the CSEs
of evolved stars. 
As noted by S. Kwok, circumstellar
environments are particularly valuable laboratories for studying chemical
synthesis and processes; the energy input from the central star is typically well
constrained, there is a high level of symmetry, and quantities
like density and temperature can be characterized as a function of
radius more readily than in many other types of environments. Further,
we have the added dimension of constraints on the evolutionary
timescales involved.

K. Menten provided a comprehensive overview of the study of the chemistry of the
envelopes of AGB stars and supergiants, highlighting some of the many
advances in this subject expected to emerge thanks to the combination
of bandwidth, sensitivity, spatial resolution afforded by new facilities,
particularly ALMA and the JVLA. 
As noted by Menten, 
one of the greatest contributions anticipated from ALMA for the study of evolved
stars will be the opportunity for the first time to study their {\em entire}
molecular envelopes. In particular, ALMA's  ability to
study molecules using imaging observations at widely
different spatial resolutions will be critical for
disentangling chemistry and excitation conditions throughout the
envelope and building up a picture of its evolution as a function of 
radius. While the SMA has already made some important
contributions in this area, it is able to observe mostly 
high-excitation lines, whereas ALMA will additionally provide access to
low-excitation lines. This will permit the study of the
composition of the molecular photosphere, element
depletion during dust formation, the acceleration of the envelope, and
the photochemistry of the outer envelope. Importantly, the angular resolution of ALMA
($<$\as{0}{5}) will allow for the first time studies of the chemistry
of the inner envelopes (including the molecule formation zones) 
on scales previously only accessible
to maser observations. 

Menten drew attention to the surprising detection of NH$_{3}$ in
the CSEs of several AGB stars and supergiants (both O-rich and C-rich),
with abundances
exceeding those predicted by chemical models
by many orders of magnitude (Menten et al. 2010). The wide bandwidth
capabilities of the JVLA, which will allow simultaneous imaging of ten different
NH$_{3}$ transitions, 
should be instrumental in gaining new insight into this puzzle by
providing far better constraints on models.

Advanced new software tools will play a crucial supporting role in
the interpretation of new sub-mm data sets. 
W. Vlemmings presented a poster advertising ARTIST (Adaptable Radiative Transfer Innovations
for Submillimeter Telescopes), a next generation model suite that
allows multi-dimensional radiative transfer calculations of dust and
line emission and their polarization properties.

Finally, Kwok pointed out that expanded access to the sub-mm spectrum will
permit the study of many potentially important molecular lines that
have previously received little attention in the study of CSE
chemistry. Examples include O$_{2}$
and molecular ions.

\subsubsection{Line Surveys\protect\label{linesurveys}}
Thanks to technological advances, the last several years have seen a
rapid growth in spectral line imaging surveys---a trend that is only
expected to grow as the full capability of instruments like ALMA and
the JVLA come on line and the body of supporting
laboratory work grows. As noted by M. Claussen (National Radio
Astronomy Observatory), 
spectroscopic imaging surveys bring a number of advantages for the
study of CSEs
compared with single-dish line surveys. These include the ability
to study molecular/isotopic abundances as a function of radius, 
to infer excitation temperature of molecules based on their location
in the CSE, to identify molecules linked with dust formation,
to probe physical conditions as  a function of radius (by
looking at different transitions of the same molecule), and 
to search for complex 
structures in the line emission distribution (e.g., \S\ref{massloss}).

\begin{figure*}
\centering
\scalebox{0.7}{\rotatebox{-90}{\includegraphics{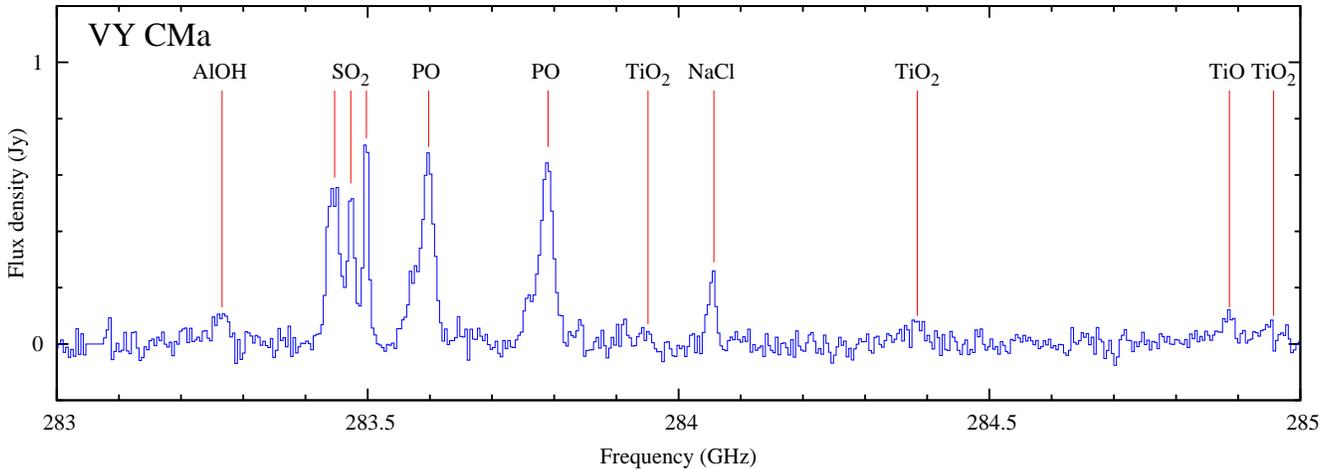}}}
\caption{Detail of a portion of the 279-355~GHz spectrum of the red supergiant 
VY~CMa obtained with the SMA by Kami\'nski et al. 2013.}
\label{fig:VYCMaspectrum}
\end{figure*}

Claussen described an ongoing line survey of 
the nearby carbon star IRC+10216 in the 18-26.5~GHz
and 18-50~GHz ranges being undertaken in conjunction with
JVLA commissioning efforts.   Despite the large number of previous single-dish
line surveys of IRC+10216, the 18-50~GHz band had never before
been surveyed. The JVLA study complements the recent mm/sub-mm surveys of this
star by Patel et al. (2009, 2011) in the 294-355~GHz range using the SMA, which detected and
imaged more than 400 lines with 3$''$ resolution. 
Importantly, the JVLA provides access to lower $J$ rotational
transitions of hydrocarbon molecules. 

As described by K. Young (Harvard-Smithsonian Center for Astrophysics),
one of the most important results from the Patel et al. survey was
the discovery of a significant population of narrow line features
(FWHM$\sim$8~\kms) having widths significantly smaller than twice the
outflow velocity from the star ($2V_{\rm
  exp}\approx$30~\kms). A significant fraction of these
narrow lines arose from unidentified species, although most are suspected to be
vibrational lines of common molecules like HCN, SiS, and SiO$_{2}$ for
which the vibrational spectrum has not yet been measured (or not yet
been made available). Young presented results from a new survey
aimed at better characterizing these narrow lines using high-resolution
($\sim$\as{0}{3})
observations with the e-SMA (i.e, the SMA in combination with the
Caltech Submillimeter Observatory and the James Clerk Maxwell
Telescope). The survey employed a tuning that
covered 31 of the narrow lines detected by Patel et al. Twenty-one of these
were detected with the e-SMA while the emission regions giving rise to the
remainder appear to have been
resolved out. 

A key result of the new e-SMA survey of IRC+10216 was the finding 
that the velocity widths of the majority of the detected narrow lines
increase with increasing distance from the central star. Furthermore,
Young showed that the trend seen in
the change in line profile shape as a function of radius is 
inconsistent with a constant acceleration
model. A linear acceleration model provides an improved fit, although
at present the signal-to-noise ratio of the data is insufficient to
distinguish between various more complex acceleration profiles.

Young stressed that analyses of this type are particularly important in the case of
carbon-rich stars, since these stars lack the strong maser transitions of
oxygen-rich stars (that more readily enable sampling the acceleration
region of their envelopes). The acceleration zones of carbon-rich versus
oxygen-rich 
stars may also differ owing to the different chemistries of
their dust. New mm/sub-mm interferometers are expected to provide  a
boon to this type of study by enabling resolution of the acceleration
zones of many nearby evolved stars.

To date, only about one-fourth of the molecular species identified in
circumstellar environments have been seen toward 
oxygen-rich stars. Furthermore, until now, the only
dedicated, unbiased line survey of an evolved star targeted 
carbon-rich IRC+10216. However, two workshop speakers presented
results from new spectral surveys of oxygen-rich stars (one AGB star
and one supergiant).

T. Kami\'nski (Max Planck Institut f\"ur Radioastronomie) 
described the ongoing analysis of line survey
data in the 279-355~GHz range obtained with the SMA on the peculiar
oxygen-rich RSG VY~CMa,
one of the largest stars in the Galaxy (Kami\'nski et al. 2013;
Gottlieb et al., in preparation).  
He reported the detection of a total of 210
spectral features,  all but 15 of which have been
identified. Roughly half of the lines in the spectrum are sulfur
compounds, a signature of shocks. Additionally, both TiO and TiO$_{2}$
were detected (Figure~\ref{fig:VYCMaspectrum}). 
TiO is extremely rare in the CSEs of evolved stars and
has never before been detected in the radio, while TiO$_{2}$ has never been
previously detected in an astronomical source (Kami\'nski et
al. 2013). 
These molecules have
been suspected as being important in dust formation in oxygen-rich
evolved stars. A problem, however, is that in VY~CMa the temperatures
derived for both TiO and TiO$_{2}$ are far lower than model
predictions. One solution could be that dust is forming in the hotter
inner envelope, while the observed the molecules are the products
of dust {\em destruction} (e.g., due to convective activity and
shocks). 
In the future, the higher resolution of ALMA should be able to
provide insight into this question.

A presentation by E. De~Beck (Max Planck Institut f\"ur Radioastronomie)
(delivered in absentia by M. Maercker) described
results of an SMA survey of the oxygen-rich AGB star IK~Tau in the
same 279-355~GHz range as the IRC+10216 survey described by
Kami\'nski. Approximately 200 lines have been identified. These
include lines from molecules such as PN that have never before
been seen in O-rich AGB stars. Also seen are
dust-related species such as AlO, sulfur-bearing molecules (possibly
indicating shocks), and multiple isotopologues of several species which
are expected to be vital in constraining  nucleosynthesis inside the star.

The results described by Claussen, Young, Kami\'nski, and De~Beck 
underscore the importance of expanding
unbiased line surveys (particularly line imaging surveys) 
to evolved stars spanning a broader range of
chemistries and mass-loss rates. However, it was pointed out that a
challenge to the analysis and interpretation of such data sets will be
accurate {\em continuum subtraction}. 
As illustrated by K. Menten, regions of the
spectrum that once looked like ``continuum'' are now seen
to be rife with weak lines when observed with sensitive new
instruments (e.g., Figure~\ref{fig:VYCMaspectrum}). 
S. Kwok pointed out that this may 
limit our ability to capitalize on improvements in sensitivity to
find new, more complex molecules. While this is less of a problem in
the JVLA bands compared with in the mm/sub-mm, accurate continuum
subtraction is also critical in the former case since, as noted
by Claussen, the continuum can otherwise be bright
enough to render many weak lines invisible.

\section{Radio Stars as Tools for Measuring the Scale of the Milky
  Way}
\subsection{The Distance to the Pleiades}
The Pleiades open cluster has long served as a benchmark for the
determination of the distance to stellar clusters through main sequence
fitting. As such, it  serves as a critical first step in the cosmological
distance ladder. However, the {\it Hipparcos} distance to the cluster is $\sim$10\%
smaller than that derived from other methods, leading to uncertainties
of $\sim$0.2 magnitude in defining its Zero Age Main Sequence, and
consequently throwing into flux theoretical models of young stars.
C. Melis (University of California, San Diego) 
described an ongoing program to resolve this discrepancy using 
VLBI astrometric measurements of Pleiades members (Melis et al. 2012). 

Melis and
colleagues identified a sample of ten
radio-bright stars in the Pleiades through a
targeted survey of its most X-ray-bright members
($L_{X}\gsim10^{30}$~erg~s$^{-1}$). Most, if not all, of the targets so
selected have turned out to be binaries. The target stars are weak ($\sim100\mu$Jy)
requiring use of a high sensitivity array comprising the VLBA, the
Green Bank Telescope (GBT), Effelsberg, and Arecibo, and 900 hours of
observations over the course of two years will be needed to complete
this study.  The outcome will be absolute parallaxes
for ten Pleiades members with 1\% precision. One indirect outcome may also
be clues as to how the presence of a companion affects the
radio luminosity of young stars. A possibility suggested by Melis is
that the companion disrupts the tidal locking to the disk during the
protostellar phase, preventing the stars from being spun down.

\subsection{Mapping the Structure of the Milky Way}
Our vantage point within the Milky Way Galaxy has long posed a
severe challenge to astronomers attempting to measure its
size scale and spiral structure.
M. Reid 
described how new VLBI measurements of masers from young massive stars, compact 
\HII\ regions, and evolved supergiants are
providing fundamental new insights into these questions. 

Reid pointed out that thanks to current calibration techniques,
the accuracy of parallax measurements from VLBI
techniques ($\sim10\mu$as) already meets or exceeds that expected
from the {\it GAIA} optical astrometric space mission. A further 
unique advantage is that only
with radio wavelength VLBI techniques can such measurements be made in
the plane of the Milky Way where extinction is severe, thus permitting
measurements of the spiral structure.

\begin{figure}
\centering
\scalebox{0.6}{\rotatebox{0}{\includegraphics{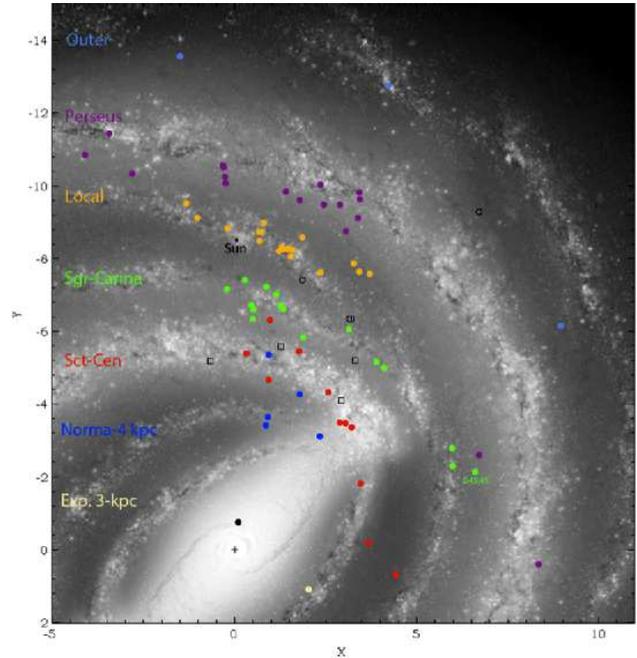}}}
\caption{Locations of massive star forming regions with H$_{2}$O or
  CH$_{3}$OH maser parallaxes measured by the VLBA (the BeSSeL Survey), VERA, and the
  EVN, overlaid on an artist's conception of the Milky Way from R. Hurt
  (NASA/JPL-Caltech/SSC).  Assignments to different spiral arms are based on CO
  longitude-velocity data and are indicated by the different colors.
  Image courtesy of M. Reid. }
\label{fig:spiralarms}
\end{figure}

Reid described recent results from the ongoing BeSSeL program, a VLBA key
project to map the Milky Way's structure. This program (with
Reid as Principal Investigator) was
awarded more than 5000 observing hours over a five year period to
accomplish this goal through the measurement of hundreds of
parallaxes and proper motions of H$_{2}$O and CH$_{3}$OH
masers. Observing began in 2010, and $\sim$70 sources have been
observed so far, yielding $\sim$60 parallax measurements. 
Measurements are also being obtained with the European VLBI Network (EVN) and
the VLBI Exploration of Radio Astronomy (VERA) array.

Results to date are consistent with the Milky Way having four
major gaseous spiral arms plus additional minor ones near its bar
Figure~\ref{fig:spiralarms}). 
The data permit characterization of the pitch angle of these arms,
yielding values of $\sim13^{\circ}$ for the outer arms and slightly 
smaller values for the inner arms. The data have
also enabled the derivation of a new azimuthally averaged Milky Way
rotation curve. The new curve is obtained by fitting the astrometric measurements to a
model of the Galaxy. Unlike previous derivations that utilized only (1-D)
radial velocities, it takes advantage of the newly measured
3-D motions and the new
``gold standard'' distances (e.g., Reid et al. 2012).

The analysis by Reid and collaborators has led to a new, accurate
distance to the Galactic Center $R_{0}=8.38\pm0.18$~kpc and a circular
rotation speed for the Sun of $\Theta_{0}=243\pm7$~\kms. These results
clearly indicate the need for a revision in the International Astronomical
Union (IAU) recommended
values for these parameters.

\section{White Dwarf Binaries}
J. Sokoloski reviewed the properties of accreting white dwarf binaries
including cataclysmic variables (CVs) and
symbiotic systems. In both types of binary, the main source
of radio emission  is outflows
(collimated, uncollimated, and/or eruptive) which give rise to
thermal bremsstrahlung and 
synchrotron emission from particles accelerated in shocks. 
Sokoloski pointed out that the recent confirmation of jets in
CVs implies that all classes of accreting WD binaries are now known to
have jets, although presently, only a handful of CV jets are known. 
In the case of symbiotics, $>$5\% show evidence for
collimated outflows, but the true fraction may be much higher, as the
jets are transient, and to
date, only a small fraction of such systems have been observed in a manner that would
have permitted identification of the presence of a jet (e.g., via high resolution
radio imaging) . 
One puzzle is why the jet fraction in CVs appears to be
so much smaller than in symbiotics; explanations may include
the larger accretion disks or larger accretion rates of symbiotics.
Regardless, Sokoloski noted that 
the fact that jets are seen in different classes of WD binaries under a wide
range of situations (e.g., various binary separations, different
classes of donor stars; with or without shell burning)
implies that the mechanism for jet formation must be quite
robust. Sokoloski also noted that in
some cases, evidence for a link between behavior in the disk and the
jet has now been confirmed (similar to what is seen in X-ray binaries).

\subsection{Novae\protect\label{novae}}
Accreting WD binaries produce the most common type of stellar
explosion in the universe: novae. 
As described by J. Sokoloski, observations of novae in the radio are key to
understanding the physics of these objects and in particular, to obtaining
more accurate estimates of the mass of their ejecta. Indeed,
one of the outstanding puzzles concerning
classical novae  is an order of magnitude discrepancy between the
amount of material ejected by these explosions as inferred through
observations compared with the amount predicted by theory. Since the
ejecta mass is tied to all of the fundamental properties of the
explosion as well as the evolution of the white dwarf, 
resolution of this discrepancy is deemed of critical importance. Sokoloski
stressed that the ejecta
masses derived from observations depend on a variety of assumptions
(e.g., inherent sphericity; the shape of the density profile; 
a ``Hubble flow'' velocity profile; a fixed electron temperature), 
but that vastly improved constraints on these quantities are now
becoming possible thanks to multi-frequency imaging observations of
novae using the JVLA, including work by the EVLA Nova (eNova) 
project led by Sokoloski. 

Sokoloski described how recent work by the eNova project has also led
to some surprises.
In the current standard picture, the radio emission from a nova is initially optically
thick at all radio frequencies, and the radio flux increases with time as
the optically thick ejecta grows in size. Once the growing ball of
gas gets large enough, its density then drops sufficiently for it to
become optically thin at higher frequencies, leading to a downturn in
the radio light curve at successively lower frequencies. 
By monitoring the light curve at a wide range
of radio frequencies over time, it is thus possible to 
probe the entire temperature and density structure of the
remnant and derive a mass estimate, in a manner that can be equated to
``peeling an onion''. However, the simple picture of a homogeneously expanding shell of
ionized gas (the so-called 
Hubble flow model) breaks down for two novae recently observed with
the JVLA.

In the case of V1723~Aql, recent JVLA light curves by Sokoloski and
collaborators unexpectedly
revealed a strong deviation from model predictions, suggesting the
presence of a large quantity of optically thin material outside of the
optically thick fireball (Krauss et al. 2011). Also, in the archetype of
recurrent novae, T~Pyx, the radio emission associated with a recent
outburst was observed to be delayed by more than 50 days from the
initial outburst, indicating a multi-stage ejection in which the bulk
of the mass ($>4\times10^{-5}~M_{\odot}$---an order of magnitude higher
than theoretical predictions) was delayed 
beyond the initial eruption (Nelson et al. 2012). 

One particularly surprising development in the area of nova research came in 2010
with the discovery of $\gamma$-ray emission from the classical
nova V407~Cyg.
As related by M. Rupen (National Radio Astronomy Observatory), 
two additional $\gamma$-ray novae were detected during 2012. Our understanding of this phenomenon is
still in its infancy, but as described by Rupen, radio observations
are providing some key insights and serve as a
crucial complement to optical and X-ray measurements.

\begin{figure}
\centering
\scalebox{6.0}{\rotatebox{0}{\includegraphics{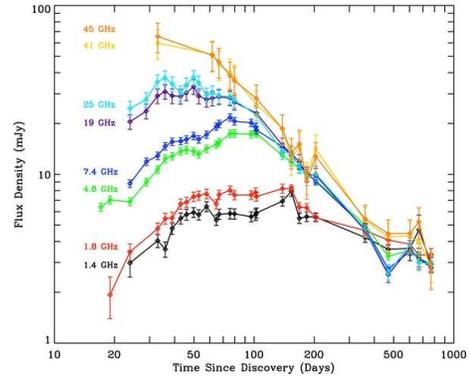}}}
\caption{Multi-frequency radio light curve for nova V407~Cyg derived
  from JVLA measurements (Chomiuk et al. 2012a). 
The optical discovery of the nova occurred on
2010 March 10.8. Reproduced by persmission of the AAS. }
\label{fig:novalightcurve}
\end{figure}

V407~Cyg is a symbiotic binary, and
JVLA observations of its nova outburst by Rupen and colleagues
(Chomiuk et al. 2012a) revealed that
the radio flux does not rise as steeply with time as predicted and
is inconsistent with an optically thick thermal source 
(Figure~\ref{fig:novalightcurve}). This
can be explained if the radio emission is dominated not by the nova
ejecta, but by a thermal
component produced as the nova blast ionizes the red giant's
wind. This is the first ever case where the radio light curve of a
nova has been seen to be dominated by this mechanism. 
As detailed by Rupen, observations with MERLIN from days 11-20 following the outburst 
(Mioduszewski et al.,
in preparation) revealed a further surprise: material extended on scales of
more than 1000~AU. Shocks/ejecta are unable to explain material at
these distances over such short timescales, implying this must be
pre-existing circumbinary material that has been flash-ionized. JVLA observations one
year later subsequently showed two components (one thermal, one synchrotron)
corresponding in position to the emission seen with MERLIN. This is
interpreted as new material colliding with the large-scale material
previously revealed by the MERLIN data.

As described by S. Deguchi, V407~Cyg has another unique trait in
being the first classic nova found to have associated
SiO maser emission (Deguchi et al. 2011). Subsequent monitoring of the SiO masers
with Nobeyama showed that the SiO maser disappeared about 20 days
after the nova eruption but then unexpectedly reappeared. This is
interpreted as evidence that the emitting regions were
destroyed by the nova shock (traveling at a speed of $\sim$3000~\kms),
then later replenished.

Rupen explained that for V407~Cyg, it was initially
suspected that interaction between the nova ejecta and the dense wind
of the red giant primary (i.e., shocks) 
was responsible for the production of its $\gamma$-ray
emission. However, the subsequent detection of $\gamma$-rays from Nova
Sco 2012 and Nova Mon 2012, 
neither of which is known to have a
similarly dense medium present, seems 
to rule out this explanation. Additionally, in the case of Nova Sco
2012, a link with
strong shocks appears to be excluded by the lack of X-ray
emission. 
So far a common thread that can explain the
production of $\gamma$-rays in these three novae remains
elusive. 
All three have been subsequently detected
in the radio, indicating significant mass outflow, and future radio
observations are expected to help better quantify this. 

\section{Supernova Progenitors}
Despite decades of research, the progenitor stars associated with the
various subclasses of supernovae remain poorly
known. However, as described in the review talk by L. Chomiuk (National Radio Astronomy
Observatory/Michigan State University), a wealth of important new
insights has been emerging from radio wavelength studies.

The radio emission from supernovae arises from synchrotron radiation 
produced by the interaction between
the supernova shock wave and the pre-existing circumstellar material. 
As this shock wave propagates through the CSE, it effectively samples the density
of the local environment as a function of distance from the star. Thus
by studying
the radio light curves of supernovae, we can probe the
mass loss history of the progenitor and hence glean important clues as to
its nature. As described by Chomiuk, the radio light curves
accumulated over the past four decades point to a diverse range of
mass-loss histories.

Chomiuk reminded us that 
radio emission (or lack thereof) was fundamental to the
recognition of a distinction between Type Ia and Type Ib/c
supernovae. Furthermore, despite numerous efforts,
no Type Ia supernova has yet to be detected in the radio. Some of the
most recent attempts were made by Chomiuk and collaborators as part of a  
program aimed at searching
for radio emission from every Type Ia supernova within 30~Mpc and
accessible to the JVLA. The absence of radio
emission from this class of supernovae argues against symbiotic progenitor
systems (in
which a white dwarf is accreting material from a red giant donor)
since their dense environments should produce detectable levels of radio emission
(Chomiuk et al. 2012b).

SN1970G was the first supernova ever detected at radio wavelengths and
posed a significant challenge to the radio instruments of its day
(e.g., Gottesman et al. 1972). 
Chomiuk reported that in 2011, it became the supernova with the longest ever recorded
light curve (more than 40 years) when it was detected with the JVLA at
a flux density of a mere
33.7$\pm$4.3$\mu$Jy at 5~GHz (Dittman et al., in preparation). 

A new insight type~IIn supernovae came in July/August 2012 when
observers were able to watch a luminous blue variable (LBV) and former ``impostor''
supernova SN2009ip explode as a
bona fide supernova.
This is the first time that a type IIn supernova has
been definitively linked with an LBV. It also represents the first
time that we
have information on the mass loss history of a star and a historical
light curve prior to its
explosion. This will thus serve an important test case for the
interpretation of mass-loss histories inferred from subsequent radio
observations. Chomiuk reported that 
JVLA monitoring of SN2009ip is ongoing 
but that no radio
counterpart had been detected as of the time of the workshop. This
is not unexpected given that type IIn supernovae are typically associated
with dense
environments where free-free absorption can impede the
radio emission from rising to the level of detectability for several
months following the explosion.

R. Ignace (East Tennessee State University) presented new ideas on
how Faraday rotation measurements could be used as a
probe of stellar wind bubbles (and supernova shells), potentially providing
clues on the nature of the central star and its wind as well as
stellar magnetic fields (Ignace \& Pingel 2013). He presented
analytic models of wind cavities threaded by magnetic fields of
different geometries (azimuthal or split monopole) 
and the characteristic signatures that 
they would impart on the observed position angle rotation
(or rotation measure) maps. He concluded that only the azimuthal field
is likely to lead to detectable Faraday rotation signals. In this
case, a characteristic antisymmetry (sign flip) is predicted in the
rotation measure map that is completely
independent of the viewing inclination. Using an azimuthal field
model, he showed that the resulting position angle map reproduces the
main features seen in observations of the supernova remnant G296.5+10.0.

Ignace also proposed the novel possibility of studying variable Faraday rotation
from the evolving ionization front of a supernova as a means to
studying the magnetic field in the ISM. In this case, the
interstellar field will be static, but as the ionized bubble grows,
different field scale lengths will be sampled. Such studies may soon be
within the realm of possibility within very nearby galaxies using VLBI techniques.

\section{Closing Discussion}
The final scientific talk of the Radio Stars workshop was an overview
and perspectives talk presented by S. Kwok. Several highlights
from this presentation are described in the relevant sections above. 
The workshop then concluded  with an open discussion moderated by T. Bastian. 
This discussion comprised in large part musings and prognostications about
the future of radio science in general rather than problems unique to the stellar
community. 

A concern voiced by several participants related to the growth of data
volumes at rates that have already begun to challenge or exceed astronomers'
ability to handle them. It was noted, for example, that 
by 2013, the JVLA will
already be producing more than a petabyte per year. The problem becomes
particularly severe for dynamic imaging spectroscopy (see also 
\S\ref{sunprospects}), where the time,
space, frequency, and polarization information effectively produce a
5-D data set, and
it was questioned whether computing and software tools will be able
to keep pace with growing needs.  Confronting this issue was deemed vital
to insuring maximum scientific return on our investments in new
instrumentation. 

It was recognized that scientists may have to approach modern radio
data with
the intent of asking very specific questions (e.g., characterization
of a particular spectral feature) and that it will become
impossible to take advantage of the entire information content of most
data sets. Similarly,
traditional approaches to interferometric data analysis will likely need to evolve to
methods that address specialized needs (e.g., fitting models directly to
the $u$-$v$ data
rather than producing large maps). A downside noted to this approach is the
possibility that science will become less curiosity-driven and less
open to the kind of serendipity that is often so crucial to driving
scientific progress.
Such fears seem 
to be reinforced by the reality of competing for limited research funding, with
pressure felt to pursue research topics deemed
high priority by funding agencies and having
well-defined outcomes.  

There was a consensus 
that given the rich, multi-dimensional nature of the data sets
expected from current and planned radio facilities, 
essentially all data sets will effectively be ``legacy'' data with the
potential to be mined for future science well beyond the goals or needs of the
original proposal. For this reason, archiving data from new
instruments, processed in a uniform and
systematic way and without loss of information, was seen to be critical.

It was pointed out that a sensitive all-sky radio survey could be a
tremendous resource for the community and a valuable complement to
counterparts planned at optical and other wavelengths. However, it was
anticipated that resources
to carry out such a survey
are unlikely to be forthcoming in the near term. Lastly, given the well-recognized
power of 
multi-wavelength science, 
another concern that was voiced was the multiple jeopardy
incurred when attempting to
secure observing time on multiple highly oversubscribed instruments
along  with subsequent
funding support to carry out the analysis. 

While no simple or clear-cut
solutions to most of the above issues emerged from the Radio Stars workshop,
it is hoped that the discussions stimulated some new thinking that
will eventually help to 
lead to constructive solutions to these
complex issues.

\acknowledgments
 LDM gratefully
acknowledges the efforts of the Radio Stars Local Organizing Committee
(H. Johnson, K. T. Paul, and  J. Soohoo) and Scientific Organizing
Committee (M. Rupen, K. Menten, E. Humphreys, and N. Patel).
Financial support for this workshop was provided by 
a grant from the National Science Foundation (AST-1241363). 

\references

Ainsworth, R. E. et al. 2012, MNRAS, 423, 1089

Amiri, N., Vlemmings, W. H. T., Kemball, A. J., \& van Langevelde,
H. J. 2012, A\&A, 538, A136

Benz, A. O. \& G\"udel, M. 1994, A\&A, 285, 621

Bower, G., C., Plambeck, R. L., Bolatto, A., McCrady, N., Graham,
J. R., de Pater, I., Liu, M. C., Baganoff, F. K. 2003, ApJ, 598, 1140

Chau, W., Zhang, Y., Nakashima, J.-I., Deguchi, S., \& Kwok, S. 2012, ApJ, 760, 1

Chen, B., Bastian, T. S., White, S. M., Gary, D. E., Perley, R.,
Rupen, M., \& Carlson, B. 2013, ApJL, 763, L21

Choi, Y. K., Brunthaler, A., Menten, K. M., \& Reid, M. J. 2012, in
Planetary Nebule: An Eye to the Future, IAU Symposium 283,
ed. A. Manchado, L. Stanhellini, \& D. Sch\"onberner, 330

Chomiuk, L., Krauss, M. I., Rupen, M. P. et al. 2012a, ApJ, 761, 173

Chomiuk, L., Soderberg, A. M., Moe, M., Chevalier, R. A., Rupen,
M. P., Badenes, C., Margutti, R., Fransson, C., Fong, W.-f., \&
Dittmann, J. A. 2012b, ApJ, 750, 164

Cranmer, S. R. 2009, ApJ, 701, 396

Cranmer, S. R. \& Saar, S. H. 2011, ApJ, 741, 54

De Beck, E., Decin, L., de Koter, A., Justtanont, K., Verhoelst, T.,
Kemper, F., \& Menten, K. M. 2010, A\&A, 523, 18

Deguchi, S., Koike, K., Kuno, N., Matsunaga, N., Nakashima, J.-I., \&
Takahashi, S. 2011, PASJ, 63, 309

Deguchi, S., Matsunaga, N., \& Fukushi, H. 2005, PASJ, 57, L25

Feldman, P. A. \& Kwok, S. 1979, JRASC, 73, 271

Fok, T. K. T., Nakashima, J., Yung, B. H. K., Hsia, C.-H., \& Deguchi,
S. 2012, ApJ, 760, 65

Forbrich, J., Osten, R. A., \& Wolk, S. J. 2011, ApJ, 736, 25

Forbrich, J. \& Wolk, S. J. 2013, A\&A, 551, A56

G\'erard, E., Le Bertre, T., \& Libert, Y. 2011, in Proceedings of the
Annual Meeting of the French Society of Astronomy and Astrophysics,
ed. G. Alecian, K. Belkacem, R. Samadi, and D. Valls-Gabaud, 419

Girart, J. M., Patel, N., Vlemmings, W. H. T., \& Rao, R. 2012, ApJ,
751, L20

Gottesman, S. T., Broderick, J. J., Brown, R. L., Balick, B., \&
Palmer, P. 1972, ApJ, 174, 383

Gray, M. D. 2007, MNRAS, 375, 477

G\"udel, M. \& Benz, A. O. 1993, ApJ, 405, L63

Hallinan, G., Antonova, A., Doyle, J. G., Bourke, S., Lane, C., \& Golden,
A. 2008, ApJ, 684, 644

Hallinan, G., Bourke, S., Lane, C., Antonova, A., Zavala, R. T.,
Brisken, W. F., Boyle, R. P., Vrba, F. J., Doyle, J. G., \& Golden,
A. 2007, ApJ, 663, L25

Hallinan, G., Sirothia, S. K., Antonova, A., Ishwara-Chandra, C. H.,
Bourke, S., Doyles, J. G., Hartman, J., \& Golden, A. 2013, 
ApJ, 762, 34

Ignace, R. \& Pingel, N. M. 2013, ApJ,  765, 19

Kami\'nski, T. et al. 2013, A\&A, in press (arXiv:1301.4344)

Krauss, M. I., Chomiuk, L., Rupen, M., Roy, N., Mioduszewski, A. J.,
Sokoloski, J. L., Nelson, T., Mukai, K., Bode, M. F., Eyres, S. P. S.,
\& O'Brien, T. J. 2011, ApJ, 739, L6

Kuznetsov, A. A., Doyle, J. G., Yu, S., Hallinan, G., Antonova, A., \&
Golden, A. 2012, AjP, 746, 99

Leal-Ferreira, M. L., Vlemmings, W. H. T., Diamond, P. J., Kemball,
A., Amiri, N, \& Desmurs, J.-F. 2012a, in Cosmic Masers -- from OH to
H$_{0}$, IAU Symposium 287, ed. R. S. Booth, E. M. L. Humphreys, \& W. H. T. Vlemmings, 79

Leal-Ferreira, M. L., Vlemmings, W. H. T., Diamond, P. J., Kemball,
A., Amiri, N, \& Desmurs, J.-F. 2012b, A\&A, 540, A42

Le Bertre, T., Matthews, L. D., G\'erard, E., \& Libert, Y. 2012, MNRAS, 422, 3422

Linsky, J. L. \& Haisch, B. M. 1979, ApJ, 229, 27

Maercker, M. et al. 20120, Nature, 490, 232

Matthews, L. D., Le Bertre, T., G\'erard, E., \& Johnson, M. C. 2013,
AJ, in press (arXiv:1301.7429)

Matthews, L. D., Libert, Y., G\'erard, E., Le Bertre, T., Johnson,
M. C., \& Dame, T. M. 2011, AJ, 141, 60

Matthews, L. D., Marengo, M., Evans, N. R., \& Bono, G. 2012, ApJ,
744, 53

McLean, M., Berger, E., \& Reiners, A. 2012, ApJ, 746, 23

Melis, C., Reid, M. J., Mioduszewski, A. J., Stauffer, J. R., \&
Bower, G. C. 2012, IAU Symposium 289, Advancing the
Physics of Cosmic Distances, ed. R. de Grijs \& G. Bono, in press (arXiV:1211.4849)

Menten, K. M. et al. 2010, A\&A, 521, L7

Menten, K. M., Reid, M. J., Kami\'nski, T., \& Claussen, M. J. 2012,
A\&A, 543, 73

Nelson, T., Chomiuk, L., Roy, N., Sokoloski, J. L., Mukai, K., Krauss,
M. I., Mioduszewski, A. J., Rupen, M. P., \& Weston, J. 2012,
submitted to ApJ (arXiv:1211.3112)

Oberoi, D. et al. 2011, ApJ, 728, L27

Osten, R. A. et al. 2010, ApJ, 721, 785

Osten, R. A. \& Bastian, T. S. 2008, ApJ, 674, 1078

Patel, N., Young, K. H., Br\"unken, S., Wilson, R. W., Thaddeus, P.,
Menten, K. M., Reid, M., McCarthy, M. C., Dihn-V-Trung, Gottlieb,
C. A., and Hedden, A. 2009, ApJ, 692, 1205

Patel, N., Young, K. H., Gottlieb, C. A., Thaddeus, P., Wilson, R. W.,
Menten, K. M., Reid, M. J., McCarthy, M. C., Cernicho, J., He, J.,
Br\"unken, S., Trung, D.-V. \& Keto, E. 2011, ApJS, 193, 17

Reid, M. J. 2012, in Cosmic Masers - from OH to H$_{0}$, IAU Symposium
287, ed. R. S. Booth, E. M. L. Humphreys, \& W. H. T. Vlemmings, 359

Richards, A. M. S., Elitzur, M., \& Yates, J. A. 2011, A\&A, 525, 56

Richards, M. T., Waltman, E. B., Ghigo, F., D., \& Richards,
D. St. P. 2003, ApJS, 147, 337

Route, M. \& Wolszczan, A. 2012, ApJ, 747, L22

Sahai, R., Claussen, M. J., Schnee, S., Morris, M. R., \& S\'anchez
Contreras, C. 2011, ApJ, 739, L3

Sahai, R. \& Nyman, L.-\ang. 1997, ApJ, 487, L155

S\'anchez Contreras, C. \& Sahai, R. 2012, ApJS, 203, 16

Trigilio, C., Leto, P., Umana, G., Buemi, C. S., \& Leone, F. 2011,
ApJ, 739, L10

Vlemmings, W. H. T., Ramstedt, S., Rao, R., \& Maercker, M. 2012,
A\&A, 540, L3

Yung, B. H. K., Nakashima, J.-I., Imai, H., Deguchi, S., Diamond,
P. J., \& Kwok, S. 2011, ApJ, 741, 94

Zhang, B., Reid, M. J., Menten, K. M., Zheng, X. W., \& Brunthal,
A. 2012, A\&A, 544, A42

\end{document}